\documentclass[11pt]{article}

\usepackage[final]{acl}

\usepackage{times}
\usepackage{latexsym}

\usepackage[T1]{fontenc}
\usepackage[utf8]{inputenc}

\usepackage{microtype}

\usepackage{inconsolata}

\usepackage{graphicx}

\usepackage{amsmath, amssymb, bm}
\usepackage{booktabs}
\usepackage{multirow}
\usepackage[normalem]{ulem}
\useunder{\uline}{\ul}{}
\usepackage{enumitem}
\usepackage{subcaption}
\usepackage[most]{tcolorbox}
\usepackage[table]{xcolor}
\usepackage{xcolor}
\usepackage{bbm}

\def\algname{\texttt{MoRe}}

\title{One Model, Many Minds: Unlocking Multi-Agent Synergy in a Single Agent via Mixture of Roles} 

\author{
 \textbf{Zhichen Zeng\textsuperscript{1}},
 \textbf{Huiyuan Chen\textsuperscript{2}},
 \textbf{Jingru Cheng\textsuperscript{2}},
 \textbf{Juan Zha\textsuperscript{2}},
 \textbf{Ming Liu\textsuperscript{2}},
 \textbf{Ying Chen\textsuperscript{2}},
\\
 \textbf{Xiyuan Yang\textsuperscript{1}},
 \textbf{Chaosheng Dong\textsuperscript{2}},
 \textbf{Haiyang Zhang\textsuperscript{2}},
 \textbf{Hanghang Tong\textsuperscript{1}}
\\
\\
 \textsuperscript{1}University of Illinois Urbana-Champaign,
 \textsuperscript{2}Amazon
\\
 \texttt{zhichenz@illinois.edu}
}

\newcommand{\beqa}{\begin{eqnarray}}
\newcommand{\eeqa}{\end{eqnarray}}
\newcommand{\beq}{\begin{equation}}
\newcommand{\eeq}{\end{equation}}
\newcommand{\ben}{\begin{enumerate}}
\newcommand{\een}{\end{enumerate}}
\newcommand{\bit}{\begin{itemize}}
\newcommand{\eit}{\end{itemize}}
\newcommand{\bi}{\begin{itemize} \item}
\newcommand{\ei}{\end{itemize}}

\newcommand{\begindef}{\begin{Definition} \rm}
\newcommand{\beginexa}{\begin{Example} \rm}
\newcommand{\beginthe}{\begin{Theorem} \rm}
\newcommand{\beginpro}{\begin{Proposition} \rm}
\newcommand{\beginlem}{\begin{Lemma} \rm}
\newcommand{\begincon}{\begin{Conjecture} \rm}
\newcommand{\begincor}{\begin{Corollary} \rm}

\newcommand{\bluecell}[1]{\cellcolor[HTML]{CFE2F3}#1}
\newcommand{\yellowcell}[1]{\cellcolor[HTML]{FFF2CC}#1}
\newcommand{\redcell}[1]{\cellcolor[HTML]{FADBD8}#1}

\newcommand{\eat}[1]{}

\begin{document}
\maketitle
\begin{abstract}
    Specializing Large Language Models (LLMs) toward distinct abilities underpins successes ranging from personalized assistants to multi-agent systems (MAS). 
    Single-agent paradigms rely on pre-defined personas or steering vectors to induce specialization, yet they impose a single fixed specialization that fails to adapt to diverse queries.
    Conversely, MAS achieves dynamic multi-perspective problem solving by orchestrating agents with distinct text-based roles, but fusing these specializations requires multi-turn interactions that inflate context length and inference cost.
    To address these limitations, we propose \underline{M}ixture \underline{o}f \underline{R}ol\underline{e}s (\algname), which adaptively composes multiple specializations into a single steering vector for single-turn inference.
    Specifically, \algname\ learns a diversified \textit{codebook} of steering vectors, each of which encodes a latent role.
    A query-aware \textit{router} dynamically fuses the codebook into a steering vector that encompasses multiple roles.
    By steering the backbone LLM with the composed vector, \algname\ enables multi-perspective specialization in a single-agent, single-turn inference process.
    The proposed \algname\ can be efficiently trained via a three-stage SFT curriculum and GRPO post-training, while the backbone LLM remains frozen.
    Experiments across reasoning and personality benchmarks show that \algname\ outperforms single-agent baselines by 2.2\% on average, and achieves performance on par with MAS while reducing token cost by $20\times$.
\end{abstract}
\section{Introduction}

Large Language Models (LLMs) have demonstrated remarkable capabilities across diverse natural language tasks~\cite{achiam2023gpt,grattafiori2024llama,comanici2025gemini,guo2025deepseek}.
Beyond general-purpose competence, recent studies show that LLMs benefit substantially from being \textit{specialized} toward distinct behaviors, cognitive modes, or expert roles~\cite{chen2025persona,wei2026beyond,sun2025personality}.
Conditioning a model on a specialization elicits targeted expertise and reasoning styles that a generic prompt fails to evoke, powering applications from role-conditioned reasoning to personalized assistance.

The \textit{single-agent paradigm} induces LLM specialization primarily through explicit role-playing~\cite{chen2024from,shanahan2023role,shao2023character} or implicit activation steering~\cite{turner2024activation,li2023inference}.
Specifically, role-playing assigns textual role descriptions, whereas activation steering directly injects a pre-computed steering vector into the hidden layers to elicit targeted behaviors.
While computationally efficient, both approaches rely on pre-defined specializations, i.e., textual personas or pre-computed steering vectors, that enforce a \textit{single static perspective}~\cite{wei2026beyond,sun2025personality}.
Consequently, the single-agent paradigms lack the capacity for query-specific adaptivity or multi-perspective reasoning.

In contrast, \textit{multi-agent systems (MAS)} dynamically integrate multi-perspective specialization via multi-turn interactions~\cite{liang2024encouraging,du2023improving,liu2024dynamic,wu2023autogen}.
By instantiating agents with distinct roles via explicit text-based prompts across agents, agents exchange and refine their perspectives over multi-turn interactions.
Yet this superiority comes at a steep price: MAS involve multiple agents and multi-turn interactions that induces heavy computation and long context~\cite{kim2025towards,tran2026single}.

\begin{figure*}
    \centering
    \includegraphics[width=\linewidth]{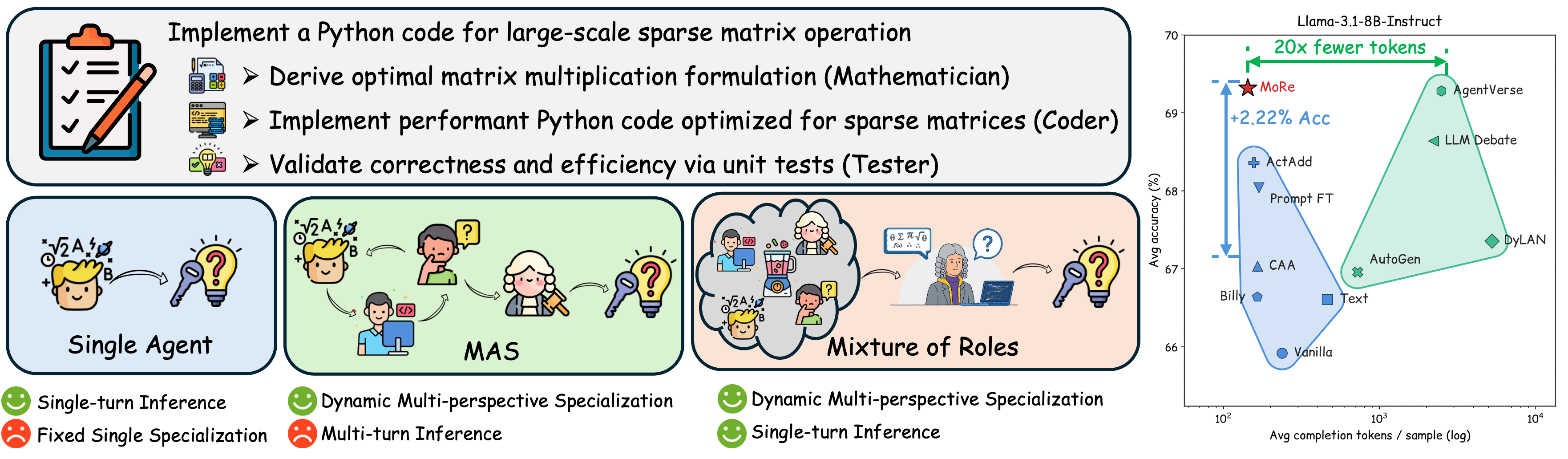}
    % \vspace{-20pt}
    \caption{Comparison between \texttt{single agent}, \texttt{MAS} and \algname. \algname\ achieves dynamic multi-perspective specialization within single-turn inference, outperforming single-agent methods while achieving higher efficiency than MAS.}
    \label{fig:teaser}
\end{figure*}

The two paradigms thus present a dilemma between computational cost and reasoning capability.
As the example shown in Figure~\ref{fig:teaser}, we are required to solve a complex task that requires mathematical derivation, algorithmic coding, and rigorous testing.
A single agent with the fixed mathematician specialization may fall short on the remaining capabilities. 
Conversely, MAS allows agents with different specializations to collaborate, but the collaboration pipeline requires multi-turn inference with high computation.
Therefore, we ask the following research question:
\begin{quote}
\textit{Can a single agent acquire dynamic multi-perspective specialization within single-turn inference?}
\end{quote}

To answer this question, we introduce Mixture of Roles (\algname) to dynamically compose multiple roles into a single steering vector for single-turn inference.
To move beyond static steering, \algname\ parameterizes steering vectors into a learnable codebook, where each entry corresponds to a candidate specialization optimized end-to-end.
A lightweight router selects and combines these candidates into a query-specific composed steering vector.
By steering with the composed vector, we equip the single agent with a dynamic multi-perspective specialization, eliminating the textual communication overhead of MAS entirely.
Our proposed \algname\ is a lightweight module that can be efficiently trained via a three-stage SFT and GRPO post-training with the backbone LLM completely frozen.

Our contributions are summarized as follows:
\begin{itemize}[noitemsep, topsep=0pt]
    \item \textbf{Framework.}
    We enable dynamic multi-perspective specialization within single-agent inference by replacing multi-turn textual coordination with activation-space composition.

    \item \textbf{Method.}
    We propose a lightweight module named \algname\ to generate query-specific steering vectors with minimal computational overhead.
    \algname\ is efficiently trained via curriculum SFT followed by GRPO post-training with the backbone LLM entirely frozen.

    \item \textbf{Evaluation.}
    Extensive experiments show that \algname\ outperforms single-agent specialization baselines by 2.2\% on average and achieves performance competitive with MAS baselines while reducing the cost cost by $20\times$.
\end{itemize}

\section{Related Works}

\subsection{Specialized Agent}
Specializing LLMs toward targeted behaviors, domain expertise, or cognitive modes has become a foundational paradigm for solving complex tasks. 
Early approaches predominantly rely on \textit{textual prompting}~\cite{shanahan2023role,shao2023character,chen2024from}, prepending hand-crafted role descriptions to condition the generation trajectory. 
Recent works have shifted from explicit prompt engineering to implicit representation engineering, where \textit{activation steering} has emerged as an efficient alternative, directly modulating the model's hidden activations at inference time. 
Foundational works~\cite{rimsky2024steering,turner2024activation,li2023inference} derive steering vectors from contrastive activations to guide the residual stream toward desired traits.
This paradigm has been extended to persona control~\cite{chen2025persona}, with recent efforts targeting persona-related neurons~\cite{sun2025personality,wei2026beyond} or attention heads~\cite{izawa2026steering}. 
More recent studies attempt to refine vector construction via preference learning~\cite{cao2024personalized} or select predefined vectors based on situational contexts~\cite{wei2026beyond}.

However, these methods share common limitations: either the specialization is pre-defined, or only a single fixed role is applied uniformly.
In contrast, \algname\ learns a codebook of steering vector and composes them per query via a lightweight router, synthesizing hybrid specializations that no single fixed vector can express.

\subsection{Multi-Agent System}
Multi-agent system (MAS) tackle complex tasks by orchestrating multiple LLM agents with distinct, prompt-defined roles. 
CAMEL~\cite{li2023camel} pioneered role-playing between communicative agents, and AutoGen~\cite{wu2023autogen} provides a general conversational programming framework for composing agents.
AgentVerse~\cite{chen2024agentverse} dynamically recruits collaborator agents and studies emergent group behaviors, and DyLAN~\cite{liu2024dynamic} adaptively selects agent teams and interaction structures on the fly.
Multi-agent debate frameworks~\cite{du2023improving,liang2024encouraging} improve factuality and reasoning by letting agents critique and revise others answers.

Across these designs, the gains stem from fusing multiple specialized perspectives, yet the fusion is realized via expensive multi-turn text communication, where inference cost grows with the number of agents and rounds.
Besides, the inflated context may degrade reasoning quality~\cite{liu2024lost}. 
In contrast, \algname\ pursues the same multi-perspective principle but in a single-turn inference: specializations live as steering directions and are fused in activation space by a query-aware router.

\section{Methodology}

\subsection{Problem Formulation}\label{sec:formulation}

Consider a frozen LLM with $L$ layers and hidden dimension $d$.
Given a query $Q = (q_1, \dots, q_T)$, where $q_i$ is the $i$-th token, LLM produces hidden states $\bm{h}^{(\ell)}_t \in \mathbb{R}^d$ for token position $t$ at layer $\ell$.
\emph{Activation steering} intervenes on the residual stream of an intermediate layer $\ell^\star$ by adding a steering vector $\bm{v} \in \mathbb{R}^d$~\cite{rimsky2024steering}:
% \vspace{-5pt}
\begin{equation} \label{eq:steer}
    \tilde{\bm{h}}^{(\ell^\star)}_t = \bm{h}^{(\ell^\star)}_t + \alpha \frac{\| \bm{h}^{(\ell^\star)}_t \|_2}{\| \bm{v} \|_2} \bm{v},
% \vspace{-5pt}
\end{equation}
where $\alpha$ is the steering strength.

To construct the steering vector, prior works~\cite{rimsky2024steering,turner2024activation} utilize \emph{contrastive pairs} that do and do not exhibit the target behavior, e.g., persona-conditioned vs.\ plain queries, and set the steering vector as the mean difference of their activations, that is
% \vspace{-5pt}
\begin{equation}\label{eq:caa}
    \bm{v} = \mathbb{E}_{Q}\left[\overline{\bm{h}}^{(\ell^\star)}(p_+\oplus Q) - \overline{\bm{h}}^{(\ell^\star)}(p_-\oplus Q)\right]
% \vspace{-5pt}
\end{equation}
where $p_+,p_-$ denote positive and negative prompts, respectively. $\overline{\bm{h}}^{(\ell^\star)}(\cdot)$ is the hidden state of layer-$\ell^\star$ mean-pooled over answer tokens.
The expectation is taken over all queries $Q$.
However, the generated vector is heuristically pre-defined and uniformly applied regardless of the input query.

To overcome these drawbacks, we propose shifting from static single steering to dynamic multi-perspective composition. 
We aim to learn a composer $\mathcal{C}_\phi$, which maps an input query $Q$ to a dynamic, query-specific steering vector $\bm{v}_{Q}$ via \emph{mixture of roles}, that is,
% \vspace{-7pt}
\begin{equation*}
\mathcal{C}\phi: Q \mapsto \bm{v}_{Q}=\sum_{n=1}^{N} w_n(Q) \cdot \bm{e}_n,
% \vspace{-7pt}
\end{equation*}
where $\{\bm{e}_n\}_{n=1}^N$ denotes a learnable codebook of latent roles, and $w_n(Q)$ is the routing weight.
This formulation enables a single frozen LLM to acquire learnable, query-adaptive, multi-perspective capabilities on the fly.

\subsection{Composer Design}\label{sec:composer}

Our proposed \algname\ includes two key modules: a specialization codebook consisting of candidate vectors, and a query-aware router that selects and fuses the candidates.

\paragraph{Specialization codebook.}
To move beyond pre-defined specialization, we maintain a learnable codebook $\{\bm{e}_n\}_{n=1}^N$, where each entry $\bm{e}_n\in\mathbb{R}^d$ is a candidate vector that encodes a specialization as a direction in the residual stream.
Rather than random initialization, we warm-start the codebook with contrastive activation directions.
Specifically, we adopt Eq.~\eqref{eq:caa} to extract $M$ different vectors as the initialization for $\{\bm{e}_n\}_{n=1}^M$.
Details on the construction of CAA prompts are provided in Appendix~\ref{app:persona}.
For the remaining $N - M$ entries, we initialize them as the convex mixtures of $\{\bm{e}_n\}_{n=1}^M$.
This anchors the codebook in behaviorally meaningful regions of activation space, which subsequent training refines end-to-end.

\paragraph{Query-aware router.}
To dynamically compose multiple roles, a router is designed score candidate vectors accordingly given the input query.
Specifically, we associate each expert $\bm{e}_n$ with a learnable probe vector $\bm{u}_n \in\mathbb{R}^d$, concatenating which formulates the probe matrix $\bm{U}\in\mathbb{R}^{N\times d}$. 
Given the token embeddings of the input query $\bm{X} = \mathrm{Emb}(Q) \in \mathbb{R}^{T \times d}$, a multi-head cross-attention module utilizes the expert probe $\bm{U}$ as the query to attend the token embeddings $\bm{X}$ that serve as the keys and values, that is
% \vspace{-3pt}
\begin{equation}\label{eq:router-attend}
    \bm{Z} = \mathrm{MHA}\bigl(\bm{U}\bm{W}^Q, \bm{X}\bm{W}^K, \bm{X}\bm{W}^V\bigr), 
    % \vspace{-3pt}
\end{equation}
where $\bm{W}^Q,\bm{W}^K,\bm{W}^V\in\mathbb{R}^{d\times d}$ are the projection matrices.
Such design enables each candidate vector to attend the query tokens, forming a query-aware representation $\bm{Z}\in\mathbb{R}^{N\times d}$ that is further processed by a MLP layer $f_\theta: \mathbb{R}^{N\times d}\to \mathbb{R}^N$ to generate a routing logit $\bm{g}=f_\theta(\bm{Z}) \in \mathbb{R}^N$.

We further adopt Top-$K$ sparse routing to generate a composed vector $\bm{v}_{Q}$, that is
% \vspace{-5pt}
\begin{equation}\label{eq:route}
    \bm{v}_{Q} = \sum_{n \in \mathcal{T}_K} w_n \bm{e}_n,~~
    w_n = \frac{\exp(g_n)}{\sum_{j \in \mathcal{T}_K} \exp(g_j)},
% \vspace{-5pt}
\end{equation}
where $\mathcal{T}_K$ is the index set of the Top-$K$ logits in $\bm{g}$. Note that for unselected experts ($n \notin \mathcal{T}_K$), the routing weight is strictly zero.
By restricting the fusion to $K\ll N$ experts, we compose a few \emph{distinct} roles per query, rather than averaging the whole codebook into an indiscriminate direction. 

Substituting the composed vector $\bm{v}_{Q}$ into Eq.~\eqref{eq:steer} gives our final steering operation:
% \vspace{-5pt}
\begin{equation*}
    \tilde{\bm{h}}^{(\ell^\star)}_t = \bm{h}^{(\ell^\star)}_t
    + \alpha\frac{\|\bm{h}^{(\ell^\star)}_t\|_2}{\|\bm{v}_{Q}\|_2}\bm{v}_{Q},
% \vspace{-5pt}
\end{equation*}
In \algname, we steer the last token of the input query, i.e., $t=T$, at the middle layer $\ell^*=\lfloor L/2\rfloor$.

\subsection{Curriculum Supervised Fine-Tuning}\label{sec:sft}

We first optimize the composer via supervised fine-tuning (SFT) with the backbone LLM frozen.
However, direct joint optimization of the codebook and router can be suboptimal.
First, an untrained router distributes noisy, near-random routing weights, causing query-specific gradients to be propagated to arbitrary candidates.
Second, an untrained codebook lacks stable and distinct semantic anchors, making it difficult for the router to identify appropriate candidates for each query.
To avoid these, we propose a three-stage curriculum for SFT training.

In \emph{stage 1}, we only optimize the codebook to ensure that each candidate is individually effective before composition.
Specifically, for each query, we uniformly sample a single candidate $\bm{e}_n$ from the codebook and optimize the selected candidate with the next-token prediction loss $\mathcal{L}_{\text{CE}}$.

In \emph{stage 2}, we freeze the codebook and train the router to select and compose candidates via Top-$K$ routing.
To prevent the router from collapsing onto a few candidates, an auxiliary load-balancing loss is adopted~\cite{fedus2022switch}, that is
% \vspace{-5pt}
\begin{equation*}
    \small
    \mathcal{L}_{\text{lb}} = N\sum_{n=1}^N f_np_n,~~\left\{
    \begin{aligned}
        &f_n=\frac{1}{BK}\sum_{b}\mathbbm{1}[n\in\mathcal{T}_K(\bm{g}_{Q_b})]\\
        &p_n=\frac{1}{B}\sum_{b}\mathrm{Softmax}(\bm{g}_{Q_b})_n
    \end{aligned}\right.
% \vspace{-5pt}
\end{equation*}
where $B$ is the batch size, and $\mathbbm{1}(n\in\mathcal{T}_K(\bm{g}_{Q}))$ is the indicator function denoting whether candidate $\bm{e}_n$ is among the Top-$K$ for query $Q$.
The stage-2 objective is $\mathcal{L}_{\text{CE}} + \lambda_{\text{lb}}\mathcal{L}_{\text{lb}}$.

In \emph{stage 3}, after both codebook and router are warmed up, we jointly optimize both, allowing the candidate vectors to adapt to the query-dependent compositions learned by the router.

\subsection{GRPO Post-Training}\label{sec:grpo}
\begin{figure}
    \centering
    \includegraphics[width=\linewidth]{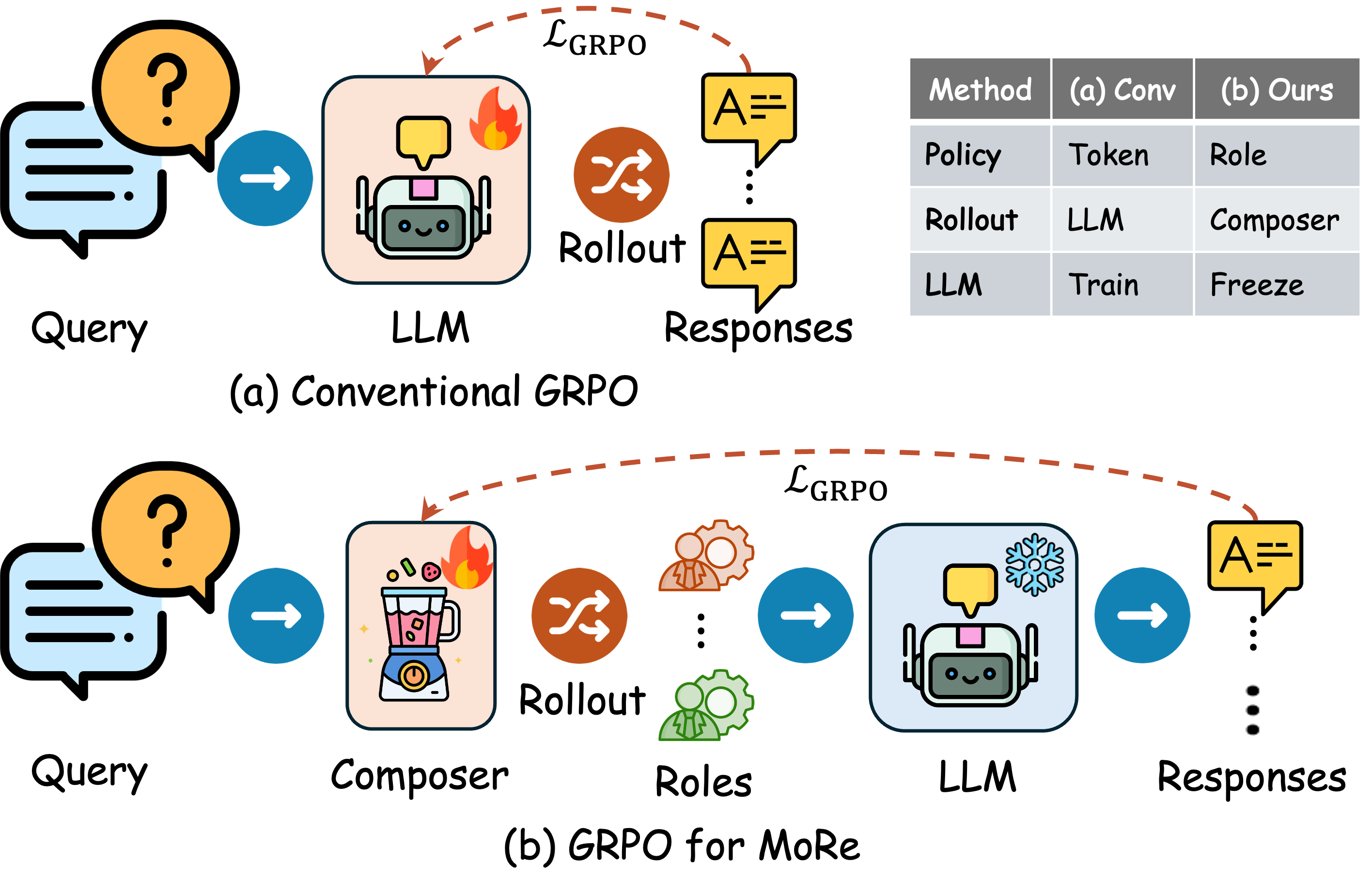}
    % \vspace{-20pt}
    \caption{Comparison between (a) conventional GRPO and (b) GRPO for \algname. Our policy is defined on the candidate role space, and the rollout only depends on the composed vector generated from the composer, making the credit assignment clear and optimization efficient.}
    \label{fig:grpo}
    % \vspace{-10pt}
\end{figure}
Although SFT equips the composer with basic routing capabilities via teacher forcing, such static imitation fails to discover optimal persona synergies.
Therefore, we utilize Group Relative Policy Optimization (GRPO)~\cite{shao2024deepseekmath} to post-train the router, encouraging the explorations of better reasoning trajectories.

Conventional GRPO optimizes an LLM's token-level policy by sampling stochastic text trajectories and reinforcing those with high group advantages.
However, this standard protocol can not be directly applied to our setting.
To train the lightweight router while keeping the backbone LLM frozen, the trajectory advantage must be \emph{attributed solely to the generated steering vector}, rather than the LLM's decoding randomness.

\paragraph{Policy.}
To decouple the reward from the LLM and fully attribute it to the router, we formulate the frozen LLM as a black-box environment.
Consequently, we shift the RL action space from the token space to the persona space.

As defined in Eq.~\eqref{eq:route}, the composed steering vector relies on the candidates and their softmax fusion.
Therefore, we define the action space as the space of all possible \emph{unordered Top-$K$-subset} $\mathcal{S} =\{S\subset\{\bm{e}_1, \dots, \bm{e}_N\}\mid |S|=K\}$.
To enable exploration for GRPO rollout, we adopt a stochastic relaxation by defining the routing policy as a stochastic mapping from query space $\mathcal{Q}$ to the action space $\mathcal{S}$, i.e., $\pi: \mathcal{Q}\mapsto \mathcal{S}$.
Instead of taking the top-$k$ logits, we \emph{randomly draw} $K$ experts based on corresponding Softmax probabilities without replacement.
As the composed vector is order-invariant, the probability of sampling a specific unordered subset $S$ given a query $Q$ marginalized over all possible permutations, that is
\begin{equation}\label{eq:policy}
    \begin{aligned}
            &\pi(S \mid Q) = \sum_{s\in\mathrm{perm}(S)}\mathrm{Pr}(s|Q)\\
            &\!\!\!\!\!=\!\!\! \sum_{s \in \mathrm{perm}(S)}\prod_{n=1}^{K}\frac{\mathrm{softmax}(\frac{\bm{g}}{\tau})_n}{\sum_{j \notin \{s_1, \dots, s_{n-1}\}} \mathrm{softmax}(\frac{\bm{g}}{\tau})_j},
    \end{aligned}
\end{equation}
where $\bm{g}$ is the routing logit, $\tau$ is the temperature, and $\mathrm{perm}(\cdot)$ is the set of all possible permutations.

Note that the policy $\pi(S \mid Q)$ is governed entirely by the lightweight router and remains independent of the backbone LLM.
Therefore, the downstream reward is uniquely attributable to the router's action, yielding precise credit assignment uncontaminated by LLM sampling noise. 
Besides, such independence offers substantial computational efficiency as rollouts are executed entirely without gradient tracking. 
GRPO only evaluates the router on the query representations, thereby eliminating the massive memory and computational overhead associated with LLM.

\paragraph{Group-relative update.}
For each query $Q$, we sample $G$ steering vector sets $\{\bm{v}_Q^j\}_{j=1}^G$, and steer the frozen LLM with each vector to produce the responses.
We adopt verifiable task reward to score each response, and collect the rewards $\{r_1, \dots, r_G\}$ for all steering vectors.
Details on reward designs are provided in Appendix~\ref{app:reward}.
With the group-normalized advantages $A_j = (r_j - \bar{r}) / (\mathrm{std}(r) + \varepsilon)$, together with the policy probability in Eq.~\eqref{eq:policy}, the router is updated by the clipped surrogate $\mathcal{L}_{\text{GRPO}}$~\cite{shao2024deepseekmath}
% \vspace{-3pt}
\begin{equation*}
\small
\begin{aligned}
&\mathcal{L}_{\mathrm{GRPO}}\!=\!-\frac{1}{G}\!\sum_{j=1}^{G}\!
\min\Bigl(\!
    \rho_j A_j,\,\!
    \operatorname{clip}(\rho_j,\!1-\epsilon,\!1+\epsilon)A_j\!
\Bigr)\!+\!\mathcal{R}_{\mathrm{KL}}\\
&\rho_j=\frac{\pi_{\mathrm{new}}(S_j\mid Q)}
         {\pi_{\mathrm{old}}(S_j\mid Q)},
\qquad
\mathcal{R}_{\mathrm{KL}}
=\beta\mathrm{KL}\!\left(
    \pi_{\mathrm{new}}\Vert\pi_{\mathrm{ref}}
\right).
\end{aligned}
\label{eq:grpo}
% \vspace{5pt}
\end{equation*}

\section{Experiments}
% \vspace{-3pt}

\subsection{Experimental Setup}\label{sec:setup}
\noindent\textbf{Models and baselines.}
We consider two LLMs: \texttt{Llama-3.1-8B-Instruct}~\cite{grattafiori2024llama} and \texttt{Qwen3-8B}~\cite{yang2025qwen3}.
We consider single agent baselines including \texttt{Vanilla LLM}, \texttt{Text Role}, Contrastive Activation Addition (\texttt{CAA})~\cite{rimsky2024steering}, \texttt{ActAdd}~\cite{turner2024activation}, \texttt{Prompt FT}~\cite{lester2021power}, \texttt{Billy}~\cite{pai2026billy}, \texttt{NPTI}~\cite{deng2025neuron}, and \texttt{IRIS}~\cite{wei2026beyond}.
We also evaluate MAS baselines including \texttt{LLM Debate}~\cite{du2023improving}, \texttt{AutoGen}~\cite{wu2023autogen}, \texttt{AgentVerse}~\cite{chen2024agentverse}, and \texttt{DyLAN}~\cite{liu2024dynamic}. More details on baselines are provided in Appendix~\ref{app:baseline}.

\begin{table*}[t]
\centering
\small
\caption{Benchmark results on reasoning tasks. We highlight \textcolor{blue}{\textbf{Top-1}}, \textcolor{red}{\ul Top-2}, and \textcolor{orange}{\textit{Top-3}} performance for each dataset.}
\vspace{-5pt}
\label{tab:reasoning}
\begin{tabular}{lccccccc}
\toprule
\textbf{Method} & \textbf{MMLU} & \textbf{TriviaQA} & \textbf{MATH} & \textbf{GSM8K} & \textbf{MedQA} & \textbf{AvgAcc} & \textbf{AvgRank} \\ \midrule
\multicolumn{8}{c}{\texttt{Llama3.1-8B-Instruct}} \\ \midrule
\texttt{Vanilla}    & 61.80 & 74.80 & 52.60 & 80.40 & 60.00 & 65.92 & 8.80 \\
\texttt{Text}       & 61.00 & 75.57 & 56.09 & 80.06 & 60.31 & 66.61 & 8.60 \\
\texttt{CAA}        & 61.34 & 73.77 & 56.31 & 82.69 & 61.06 & 67.03 & 7.60 \\
\texttt{ActAdd}     & 61.74 & 79.00 & 56.46 & 82.89 & 61.71 & 68.36 & 5.00 \\
\texttt{Prompt FT}  & 62.80 & \textit{\yellowcell 77.60} & \textit{\yellowcell 58.60} & 78.00 & \textbf{\bluecell63.20} & 68.04 & 4.60 \\
\texttt{Billy}      & 61.00 & 74.20 & 57.00 & 81.20 & 59.80 & 66.64 & 8.40 \\
\texttt{LLM Debate} & 62.20 & 74.60 & 58.40 & \textbf{\bluecell85.80} & {\ul \redcell 62.20} & \textit{\yellowcell 68.64} & \textit{\yellowcell 4.20} \\
\texttt{AutoGen}    & \textbf{\bluecell63.80} & 73.60 & 54.80 & 81.20 & 61.40 & 66.96 & 6.80 \\
\texttt{AgentVerse} & 62.40 & \textbf{\bluecell79.40} & {\ul \redcell 59.00} & {\ul \redcell 83.80} & \textit{\yellowcell 61.80} & {\ul \redcell 69.28} & {\ul \redcell 2.60} \\
\texttt{DyLAN}      & \textit{\yellowcell 63.00} & 76.00 & 54.40 & 82.40 & 61.00 & 67.36 & 6.40 \\
\algname             & {\ul \redcell 63.20} & {\ul \redcell 79.00} & \textbf{\bluecell59.20} & \textit{\yellowcell 83.40} & \textit{\yellowcell 61.80} & \textbf{\bluecell69.32} & \textbf{\bluecell2.20} \\ \midrule

\multicolumn{8}{c}{\texttt{Qwen3-8B}} \\ \midrule
\texttt{Vanilla}    & 68.80 & 61.60 & 70.20 & 88.80 & 59.20 & 69.72 & 8.40 \\
\texttt{Text}       & 67.74 & 61.43 & 70.43 & 90.11 & 59.26 & 69.79 & 7.80 \\
\texttt{CAA}        & 69.57 & 62.23 & 69.83 & 89.09 & 59.54 & 70.05 & 7.40 \\
\texttt{ActAdd}     & 69.74 & 62.63 & 70.11 & 90.00 & 60.09 & 70.51 & 6.00 \\
\texttt{Prompt FT}  & \textbf{\bluecell70.40} & 60.00 & 69.80 & 87.80 & 58.80 & 69.96 & 8.00 \\
\texttt{Billy}      & \textbf{\bluecell70.40} & \textbf{\bluecell63.40} & 67.80 & 88.20 & 60.00 & 69.96 & 5.40 \\
\texttt{LLM Debate} & 69.80 & \textit{\yellowcell 63.00} & 72.00 & \textbf{\bluecell92.40} & \textit{\yellowcell 60.80} & {\ul \redcell 71.60} & \textit{\yellowcell 3.20} \\
\texttt{AutoGen}    & \textit{\yellowcell 70.20} & 58.80 & {\ul \redcell 72.60} & 86.60 & 60.00 & 69.64 & 6.40 \\
\texttt{AgentVerse} & 69.80 & \textit{\yellowcell 63.00} & \textbf{\bluecell75.40} & {\ul \redcell 91.80} & \textbf{\bluecell62.00} & \textbf{\bluecell72.40} & \textbf{\bluecell 2.40} \\
\texttt{DyLAN}      & 69.40 & 62.00 & 70.80 & 82.20 & {\ul \redcell 61.60} & 69.20 & 6.80 \\
\algname             & \textit{\yellowcell 70.20} & {\ul \redcell 63.20} & \textit{\yellowcell 72.40} & \textit{\yellowcell 90.80} & \textit{\yellowcell 60.80} & \textit{\yellowcell 71.48} & {\ul \redcell 2.80} \\ \bottomrule
\end{tabular}
\end{table*}

\begin{table*}[t]
\small
\centering
\caption{Benchmark results on PersonalityBench. We highlight \textcolor{blue}{\textbf{Top-1}} and \textcolor{red}{\ul Top-2} performance for each dataset.}
\vspace{-5pt}
\label{tab:personality}
\begin{tabular}{lccccccc}
\toprule
\textbf{Method} & \textbf{O}           & \textbf{C}  & \textbf{E}      & \textbf{A}      & \textbf{N}        & \textbf{AvgScore}   &\textbf{AvgRank}\\ \midrule
\multicolumn{8}{c}{\texttt{Llama3.1-8B-Instruct}}                                                                        \\ \midrule
\texttt{Text}   & 8.12±1.90          & 7.06±1.25          & 9.21±0.77          & 8.15±1.68          & 9.00±1.76       & 8.31±0.85 & 4.6   \\
\texttt{CAA}    & 8.79±0.47          & 7.84±1.02          & 9.56±0.42          & 7.40±1.15          & 8.77±0.88       & 8.47±0.86 & 4.4   \\
\texttt{NPTI}   & 9.15±0.71          & {\ul \redcell 9.86±0.13}    & 9.72±0.20          & {\ul \redcell 9.29±0.40}    & 9.84±0.00       & {\ul \redcell 9.57±0.33} & 2.6   \\
\texttt{IRIS}   & {\ul \redcell 9.29±0.36}    & 9.84±0.13          & {\ul \redcell 9.76±0.27}    & 9.06±0.66          & {\ul \redcell 9.92±0.00} & 9.57±0.38 & {\ul\redcell 2.4}   \\
\algname   & \textbf{\bluecell 9.47±0.44} & \textbf{\bluecell 9.87±0.14} & \textbf{\bluecell 9.81±0.15} & \textbf{\bluecell 9.88±0.11} & \textbf{\bluecell 9.98±0.02}  & \textbf{\bluecell 9.80±0.20} & \textbf{\bluecell 1.0}\\
\midrule
\multicolumn{8}{c}{\texttt{Qwen3-8B}}                                                                                    \\ \midrule
\texttt{Text}   & {\ul \redcell 8.31±1.68}    & 6.70±1.01          & 9.39±0.44          & 8.05±1.6           & 9.12±0.94       & 8.31±1.06 & 3.4   \\
\texttt{CAA}    & 7.89±1.38          & {\ul \redcell 8.47±1.28}    & {\ul \redcell 9.47±0.93}    & {\ul \redcell 8.60±1.24}    & {\ul \redcell 9.81±0.17} & {\ul \redcell 8.85±0.78} & {\ul \redcell 2.6}   \\
\texttt{NPTI}   & 8.00±1.25          & 7.98±0.94          & 7.89±1.53          & 7.88±0.78          & 9.17±0.68       & 8.18±0.55 & 3.8   \\
\texttt{IRIS}   & 8.20±1.13          & 7.87±0.93          & 8.24±1.73          & 7.86±0.76          & 8.87±0.83       & 8.21±0.55 & 4.2   \\
\algname  & \textbf{\bluecell 8.92±1.43} & \textbf{\bluecell 9.49±0.54} & \textbf{\bluecell 9.51±0.32} & \textbf{\bluecell 9.47±0.36} & \textbf{\bluecell 9.91±0.00} & \textbf{\bluecell 9.46±0.35} & \textbf{\bluecell 1.0}\\
\bottomrule
\end{tabular}

\end{table*}

\noindent\textbf{Datasets.}
We assess task-oriented specialization on five reasoning benchmarks spanning general knowledge (\textbf{MMLU}~\cite{hendrycks2020measuring},\textbf{TriviaQA}~\cite{joshi2017triviaqa}), mathematics (\textbf{MATH}~\cite{hendrycks2021measuring},\textbf{GSM8K}~\cite{cobbe2021training}), and domain expertise (\textbf{MedQA}~\cite{jin2021disease}).
We adopt LLM-as-judge~\cite{chen2025xverify} for evaluation.
We also evaluate personality-oriented specialization on \textbf{PersonalityBench}~\cite{deng2025neuron}, covering \textbf{O}penness, \textbf{C}onscientiousness, \textbf{E}xtraversion, \textbf{A}greeableness, and \textbf{N}euroticism.
For each personality trait, methods are steered to activate (\emph{high} pole) and deactivate (\emph{low} poles) the trait on open-ended questions. 
Following the standard evaluation protocol, we adopt Claude~\cite{anthropic2024claude3} to score each response on two 1--5 scales, and report the sum and standard deviation of both \emph{high} and \emph{low} poles.

\noindent\textbf{Experiment pipeline.}
We adopt 7 candidate roles, including \textit{mathematician}, \textit{software engineer}, \textit{data scientist}, \textit{logician}, \textit{teacher}, \textit{skeptic}, and \textit{doctor}.
Details on text role design are provided in Appendix~\ref{app:exp}.
Candidate roles are utilized as the ``System Role'' for \texttt{Text}; for steering vector extraction in \texttt{CAA}, \texttt{ActAdd} and \texttt{Billy}, and codebook initialization for \algname.
For \texttt{Text}, \texttt{CAA} and \texttt{ActAdd}, we report the average performance over all roles, per-role performance is reported in Appendix~\ref{app:exp}.
All experiments are conducted on 8 NVIDIA 40G A100 GPU.

\subsection{Reasoning Results}\label{sec:exp-reasoning}
% \vspace{-3pt}
\begin{figure}[t]
    \centering
    \includegraphics[width=\linewidth]{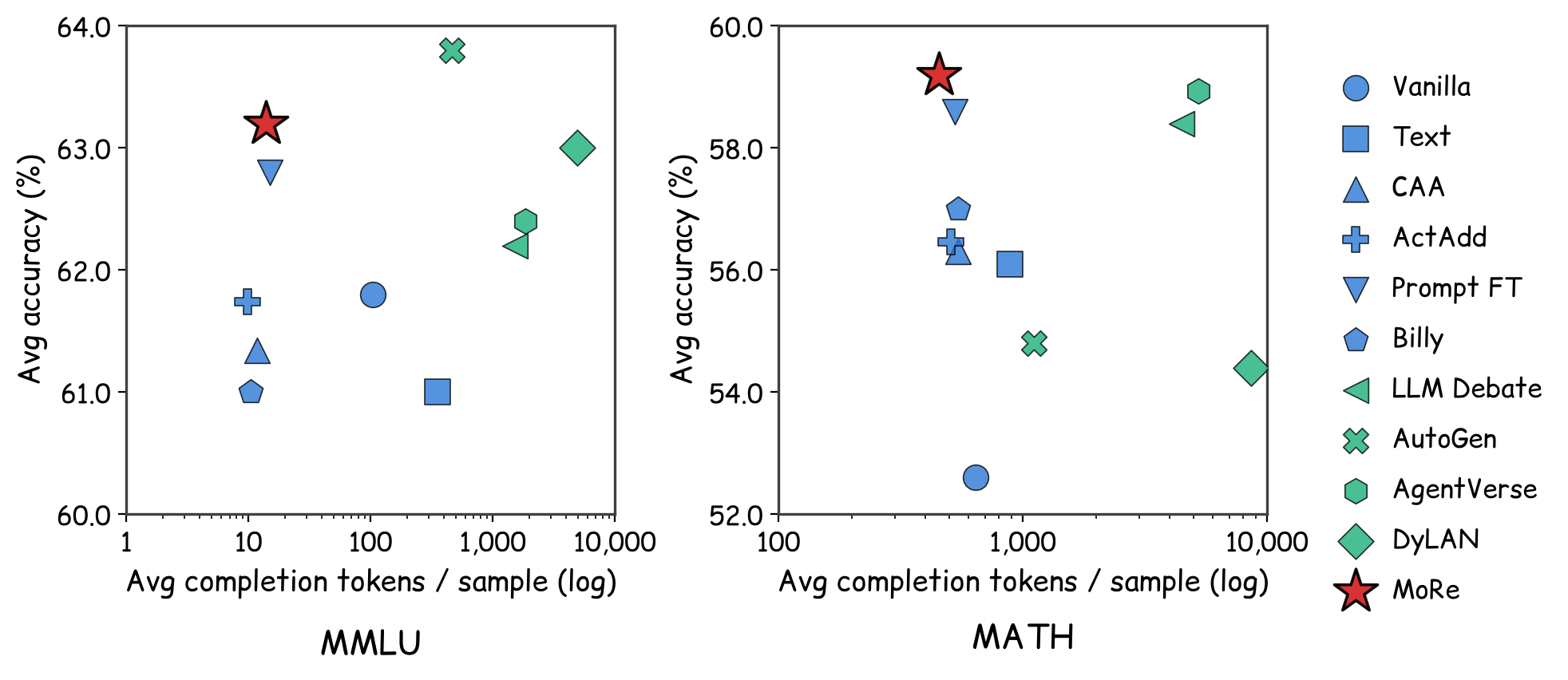}
    % \vspace{-20pt}
    \caption{Performance vs tokens. Optimal value lies at the upper left corner. \textcolor{red}{\algname} matches \textcolor{teal}{MAS} performance while only consuming tokens of \textcolor{blue}{single-agent} methods.}
    \label{fig:balance}
    % \vspace{-10pt}
\end{figure}
Table~\ref{tab:reasoning} summarizes results on reasoning benchmarks.
A comparison of model performance and token consumption is shown in Figure~\ref{fig:balance}.
Overall, \algname\ ranks Top-2 on two backbone LLMs with significantly fewer token consumption, demonstrating great balance between performance and efficiency.
Specifically, we make the following observations.

\noindent\textbf{Specialization improves LLM performance.}
Comparing with \texttt{Vanilla}, specialization helps LLM achieve higher average accuracy and better average rank.
Moreover, the per-role results in Tables~\ref{tab:app_text}--\ref{tab:app_actadd} reveal that different roles affect reasoning performance in distinct ways.
For example, \textit{mathematician} performs well on math datasets (MATH and GSM8K), but is less effective on general knowledge tasks such as TriviaQA.
These results suggest that no single role is universally optimal and that selecting appropriate specializations is important for downstream tasks.

\noindent\textbf{\algname\ outperforms single-agent baselines.}
On average, \algname\ outperforms the best single-agent baselines \texttt{ActAdd}, by 0.96 and 0.97 on Llama and Qwen, respectively.
Such improvement is more pronounced on tasks requiring longer reasoning chains.
% For example, compared with \texttt{Vanilla}, \algname\ improves MATH and GSM8K accuracy by 6.60 and 3.00 points on Llama, and by 2.20 and 2.00 points on Qwen, respectively.
Since longer reasoning involves more intermediate decisions, this provides more opportunities for specializations to influence the reasoning trajectory.
In contrast, tasks dominated by short-form recall provide fewer stages at which such specialized reasoning patterns can take effect.

\noindent\textbf{\algname\ matches MAS performance with significantly lower inference cost.}
Despite using only single-agent, single-turn inference, \algname\ achieves performance comparable to MAS, ranking the first and second on two backbone LLMs, respectively.
Unlike MAS approaches, which rely on repeated model calls, iterative discussions, and communication among multiple agents, \algname\ only requires a single model call.
As shown in Figures~\ref{fig:teaser} and~\ref{fig:balance}, \algname\ uses only approximately 1/20 as many tokens as the MAS baselines on average.
It therefore provides a favorable balancing between performance and efficiency, matching the reasoning quality of substantially more expensive multi-agent pipelines while avoiding their large token overhead.

\subsection{Personality Results}
% \vspace{-3pt}

Table~\ref{tab:personality} reports the results on PersonalityBench, covering  the Big Five personality traits.
Overall, \algname\ achieves state-of-the-art performance across backbone LLMs and different personalities.

Comparing with the best competitor, \algname\ achieves an average outperformance of 0.23 on Llama and 0.61 on Qwen, and consistently ranks the first.
In contrast, existing methods exhibit pronounced preferences across personality traits and model families.
For example, \texttt{Text} roles perform well on linguistically salient traits, e.g., Extraversion and Neuroticism, while scores lower on Conscientiousness and Agreeableness.
\texttt{IRIS} performs well on Llama (AvgRank 2.4) but degrades substantially on Qwen (AvgRank 4.2).
Such cross-model degradation suggests that the personality directions identified by existing methods may be closely tied to the internal representation geometry of a particular model.
In contrast, \algname\ consistently outperforms all baselines across both backbones and exhibits the lowest variation in average performance. 
These results demonstrate that \algname\ induces the desired personalities both effectively and consistently, with stronger robustness to model-specific representations and better generalization across personality dimensions and LLM families.

\begin{figure}[t]
    \centering
    \begin{subfigure}{.42\linewidth}
        \centering
        \includegraphics[width=\textwidth]{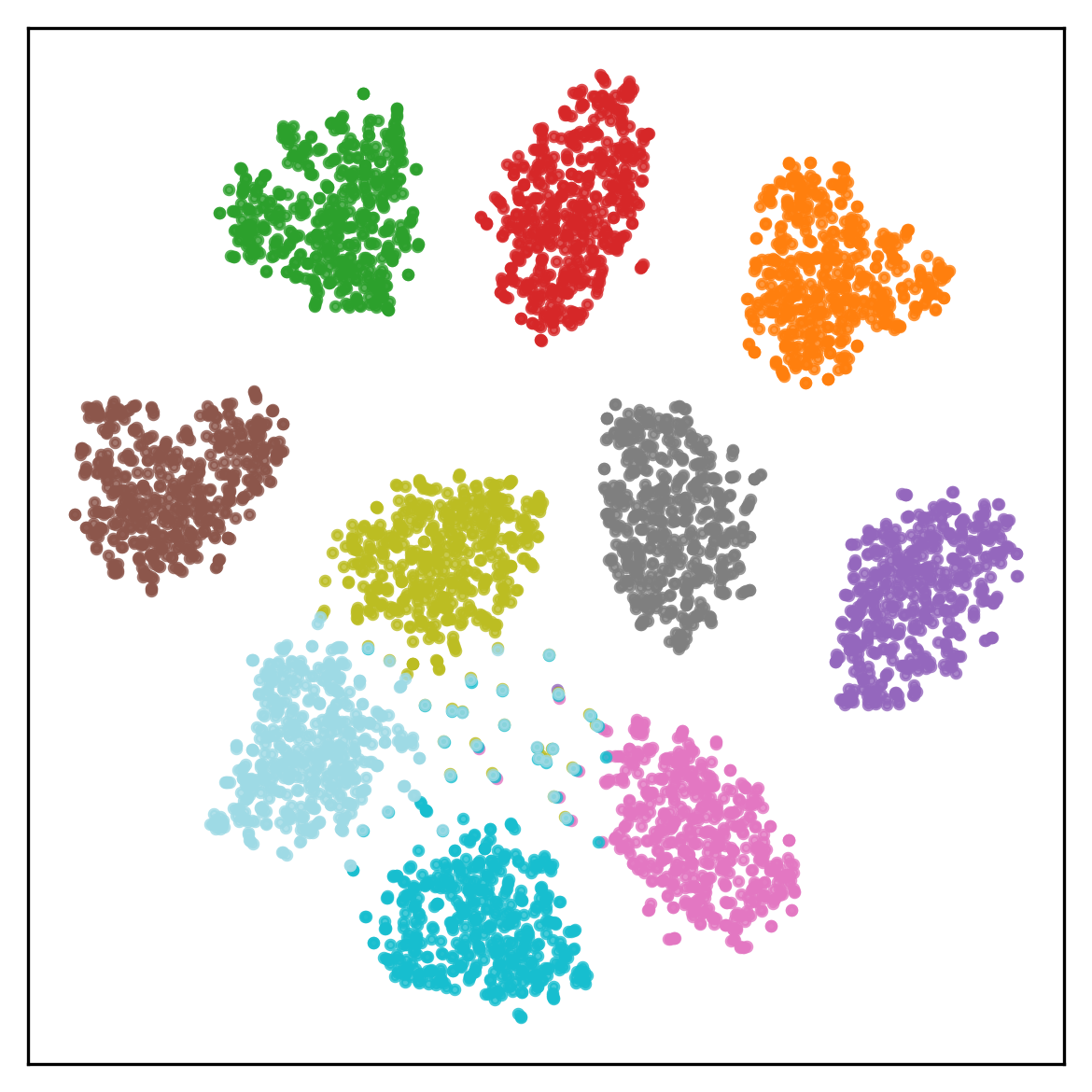}
        % \vspace{-12pt}
        \caption{Last token embedding.}
        \label{fig:codebook_1}
    \end{subfigure}
    \begin{subfigure}{.56\linewidth}
        \centering
        \includegraphics[width=\textwidth]{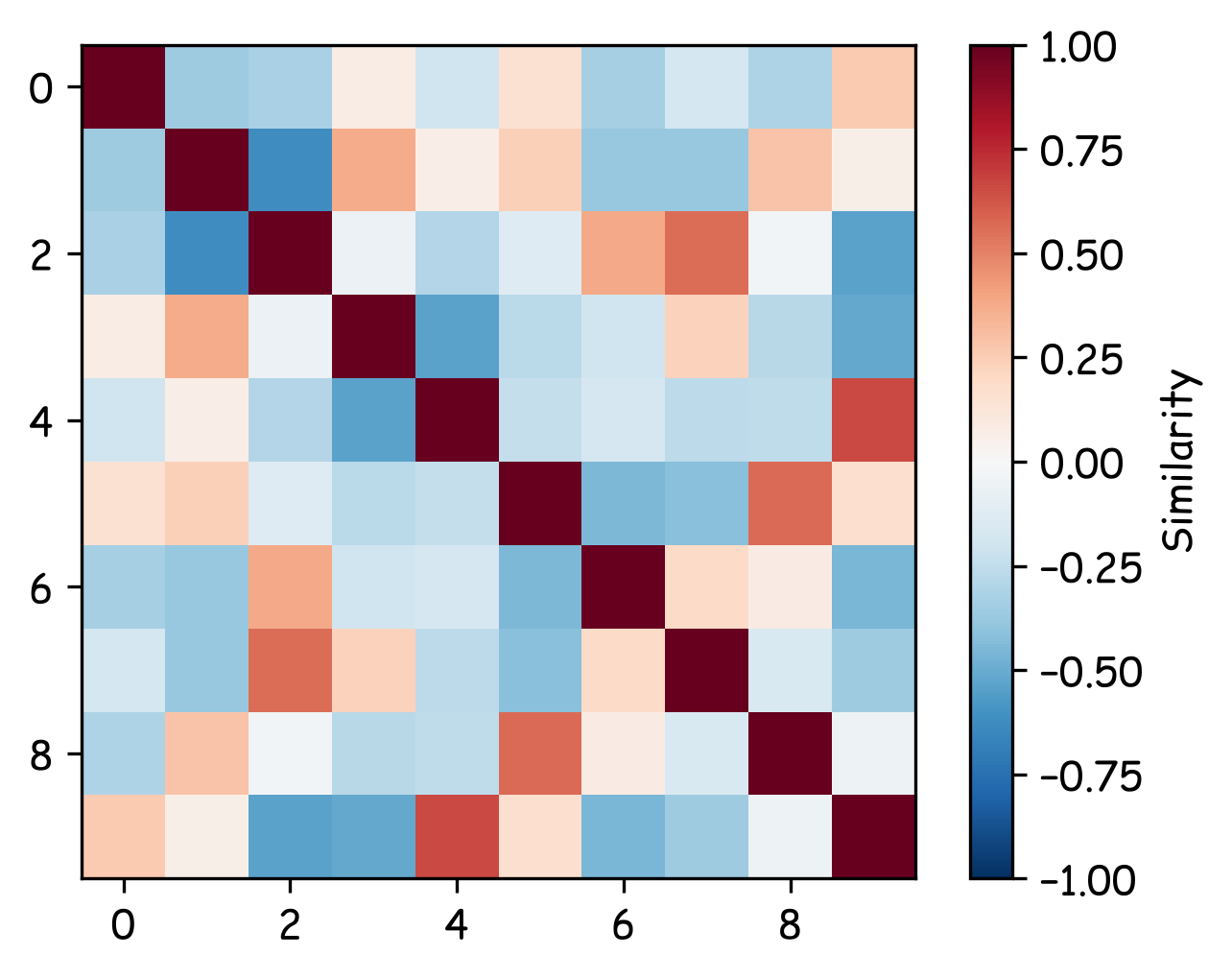}
        % \vspace{-16pt}
        \caption{Codebook similarity.}
        \label{fig:codebook_2}
    \end{subfigure}
    % \vspace{-20pt}
    \caption{Studies on the learned codebook.}
    % \vspace{-10pt}
    \label{fig:codebook}
\end{figure}
\subsection{Studies}\label{sec:exp_study}

\subsubsection{On the Learned Codebook}\label{sec:exp_codebook}

We first investigate the learned codebook.
Figure~\ref{fig:codebook_1} visualizes the hidden states of the final prompt token after applying different candidate steering vectors.
The embeddings form compact clusters, indicating that different candidate vectors redirect model representations towards distinct role-specific regions.
We further report the pairwise cosine similarities among the learned candidates in Figure~\ref{fig:codebook_2}. 
Most off-diagonal entries exhibit low similarity, showing that the candidates learn diverse steering directions even without an explicit diversity regularizer. 
This suggests that the end-to-end training objective naturally encourages functional differentiation among the latent experts. 
Moreover, the high dimensionality of the steering space provides sufficient capacity for different candidates to occupy distinct directions, thereby reducing redundancy within the codebook.

\subsection{On the Learned Router}
% \vspace{-3pt}
\begin{figure}[t]
    \centering
    \begin{subfigure}{.38\linewidth}
        \centering
        \includegraphics[width=\textwidth, trim=10 0 5 0]{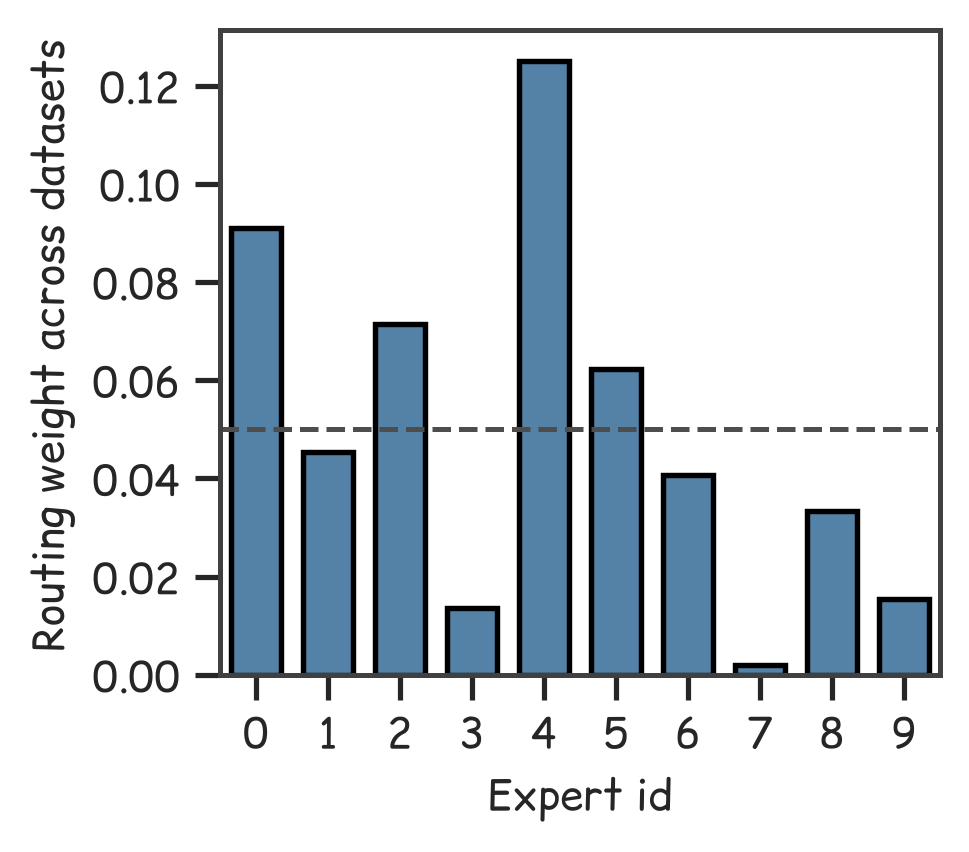}
        % \vspace{-15pt}
        \caption{Avg. routing weight}
        \label{fig:routing_1}
    \end{subfigure}
    \begin{subfigure}{.6\linewidth}
        \centering
        \includegraphics[width=\textwidth, trim=10 0 10 0]{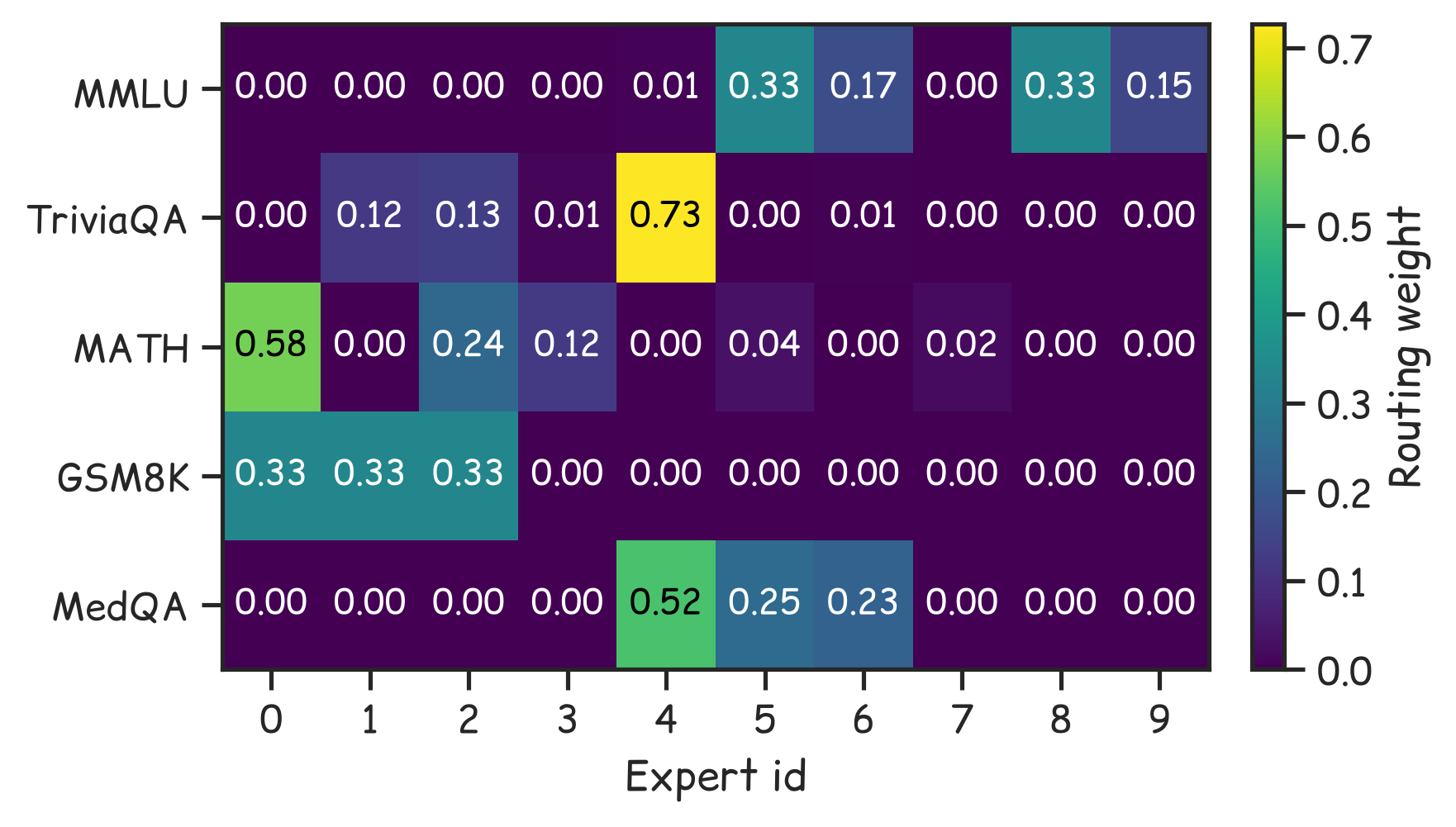}
        % \vspace{-15pt}
        \caption{Routing weight per dataset}
        \label{fig:routing_2}
    \end{subfigure}
    % \vspace{-20pt}
    \caption{Studies on the learned router.}
    % \vspace{-10pt}
    \label{fig:routing}
\end{figure}

Figure~\ref{fig:routing_1} visualizes the routing weights averaged across all datasets. 
In general, weights are distributed across most experts rather than collapsing onto a single one, while still exhibiting meaningful preferences.
In particular, Experts 0 and 4, initialized from the high-performing mathematician and teacher roles (see Table~\ref{tab:app_text}), receive noticeably higher weights, suggesting that the router effectively exploits the useful inductive biases introduced by role-based initialization.
Meanwhile, expert 7 remains nearly inactive, potentially indicating an under-utilized or overlapping specialization.

Figure~\ref{fig:routing_2} presents the routing weights for each dataset. 
Different datasets exhibit distinct routing preferences, whereas related datasets share similar patterns. 
For example, the two math benchmarks (MATH and GSM8K), heavily utilize Experts 0 initialized from the mathematician. 
MedQA instead combines Experts 4--6, initialized from the teacher, skeptic, and doctor.
In contrast, the broad disciplinary coverage of MMLU leads to a more dispersed routing pattern involving Experts 5, 6, 8, 9.
These results indicate that the router learns transferable specializations shared across related tasks and dataset-specific expert compositions, thereby adaptively matching different reasoning requirements with complementary roles.

\subsubsection{Ablation Study}
\begin{figure}[t]
    \centering
    \begin{subfigure}{.48\linewidth}
        \centering
        \includegraphics[width=\textwidth]{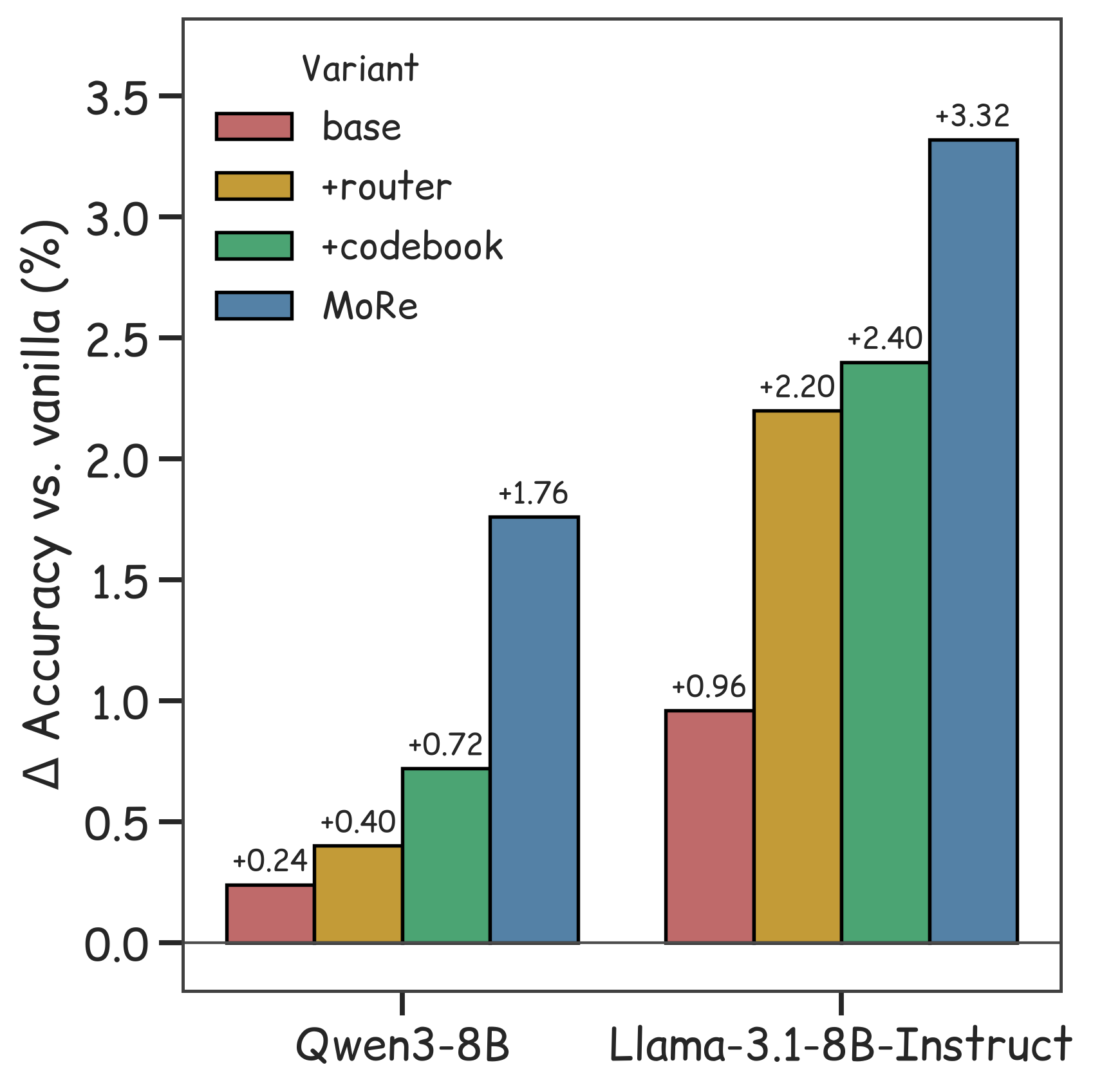}
        \caption{Router and Codebook}
        \label{fig:ablate_1}
    \end{subfigure}
    \begin{subfigure}{.48\linewidth}
        \centering
        \includegraphics[width=\textwidth]{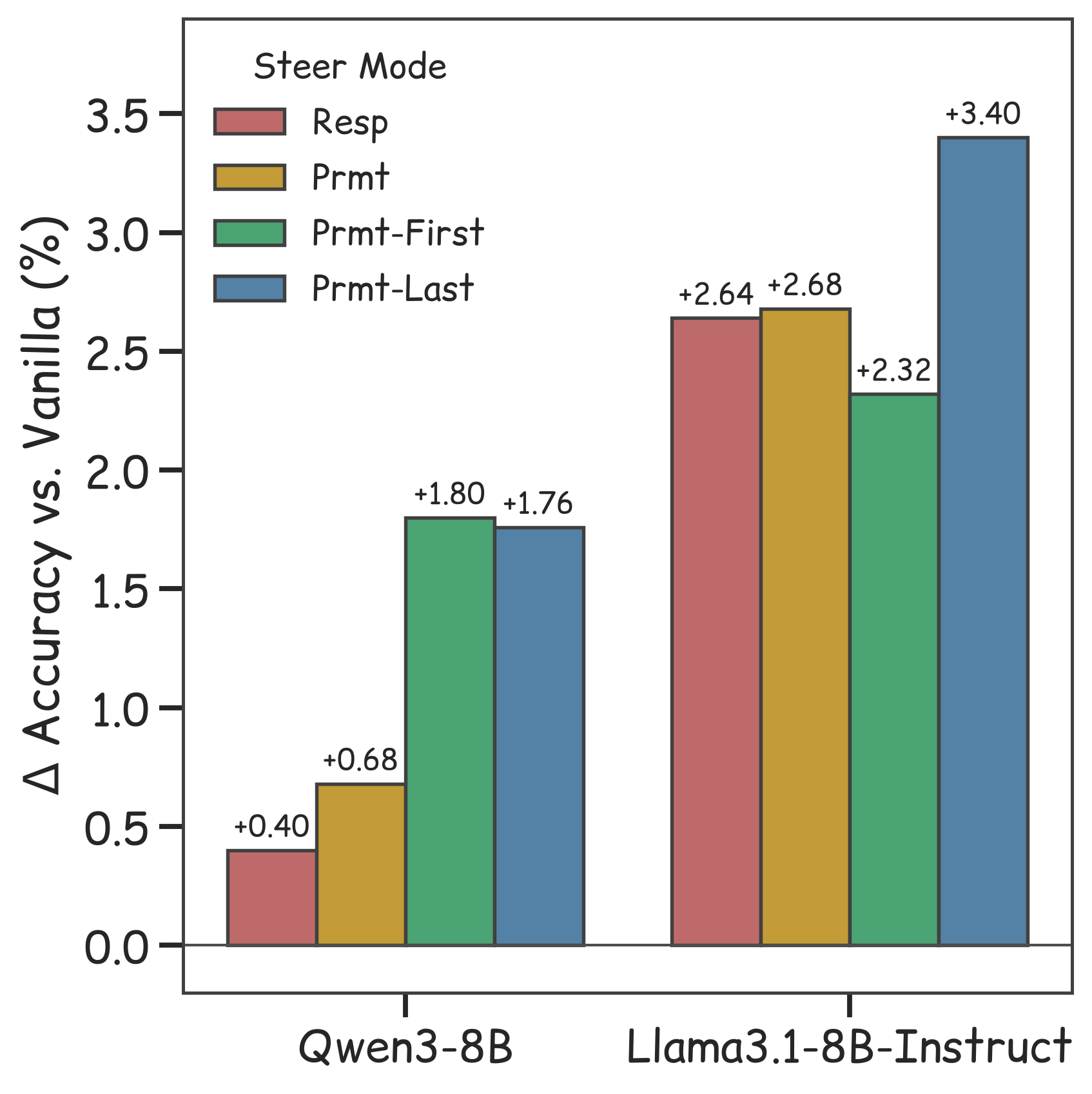}
        \caption{Steering Position}
        \label{fig:ablate_2}
    \end{subfigure}
    % \vspace{-5pt}
    \caption{Ablation studies. (a) Performance gain with \texttt{base}, \texttt{+router}, \texttt{+codebook}, and \algname; (b) Performance gain with different steering positions. Both figures use \texttt{Vanilla} performance as the baseline.}
    % \vspace{-15pt}
    \label{fig:ablation}
\end{figure}

We first study the benefits of two key modules: \textit{the learnable codebook} and \textit{the router}.
We consider four variants, including: (1) \texttt{base}, which ablates both modules and uniformly fuse fixed CAA vectors, (2) \texttt{+router}, which ablates the codebook and uses the learned router to fuse the fixed CAA vector, (3) \texttt{+codebook}, which ablates the router and uniformly fuse the learned vectors, and (4) \algname, which jointly employs both learned modules. 
The results are shown in Figure~\ref{fig:ablate_1}.
First, even uniform fusion of fixed CAA vectors (\texttt{base}) outperforms \texttt{Vanilla}, which validates the necessity of adopt multi-perspective specialization to enhance LLM reasoning.
Second, both modules contribute to the final performance: \texttt{+codebook} achieves more significant improvement than \texttt{+router}, suggesting that learning task-adaptive steering is more effective than dynamic selection among pre-defined ones.
Third, \algname\ significantly outperforms both ablated variants: the codebook provides diverse steering directions, while the router adaptively selects and combines them according to each input.

We further examine the effects of steering position using four variants, including: (1) \texttt{Resp} that steers the response tokens; (2) \texttt{Prmt} that steers all prompt tokens, (3) \texttt{Prmt-First} that steers the first prompt token, and \texttt{Prmt-Last} that steers the last prompt token.
The results are shown in Figure~\ref{fig:ablate_2}.
First, we observe that steering one single prompt token is more effective than steering all prompt or response tokens.
This suggests that a local intervention is sufficient to redirect model behaviors, whereas repeated perturbations may interfere with the original semantic representations.
Second, steering prompt is more effective than steering response, which indicates that steering before generation can better propagate throughout the entire reasoning process. 
Third, the optimal position to steer is the final prompt token, as it summarizes the preceding context and serves as the starting point for subsequent generation.
Therefore, the final prompt token acts as an effective bridging point for controlling the subsequent reasoning trajectory.

\subsubsection{Hyperparameter Study}\label{sec:hyper}
\begin{figure}[t]
    \centering
    \includegraphics[width=\linewidth, trim=10 0 10 0]{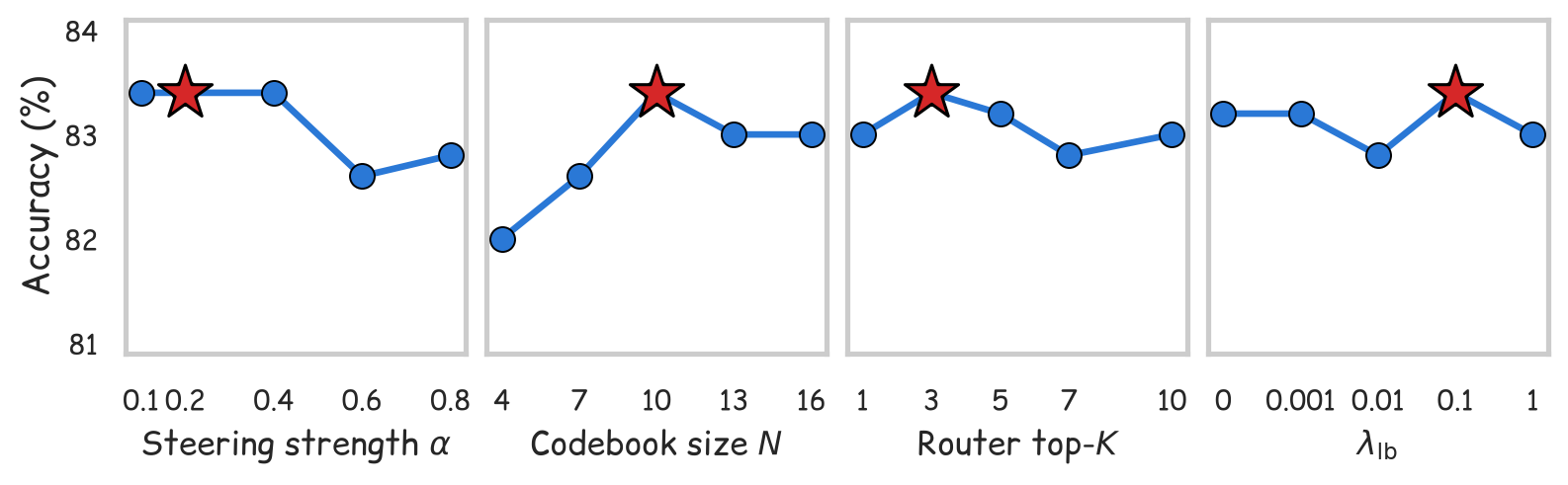}
    % \vspace{-20pt}
    \caption{Hyperparameter study on Llama+GSM8K.}
    \label{fig:hparam}
    % \vspace{-15pt}
\end{figure}

Figure~\ref{fig:hparam} studies the sensitivity of \algname\ to the steering strength $\alpha$, codebook size $N$, router top-$K$, and load-balancing weight $\lambda_{\mathrm{lb}}$.
Overall, \algname\ remains stable across a broad range of configurations.
For the steering strength $\alpha$, a small $\alpha$ is sufficient to redirect model behavior, whereas an excessively strong $\alpha$ over-perturbs the original representations.
Similarly, a moderate codebook size $N$ provides sufficient diversity, while an overly small codebook limits its diversity and a larger one introduces redundancy.
Moreover, we observe that a moderate routing $K$ generally performs well: too sparse routing (e.g., Top-1) loses the power of multi-perspective fusion, while too dense routing (e.g., Top-10) fails to achieve adaptivity to queries hence diluting expert specialization.
Finally, a moderate $\lambda_{\text{lb}}$ yields the best result, suggesting a favorable trade-off between preventing routing collapse and preserving expert specialization.
% \vspace{-3pt}
\section{Conclusion}\label{sec:conclusion}
% \vspace{-3pt}

In this work, we introduced \algname, a lightweight module to enable dynamic multi-perspective specialization within a single-agent, single-turn inference process.
\algname\ learns a diversified codebook of specializations and employs a router to adaptively compose them into a multi-perspective steering vector for each input.
This design combines the efficiency of single-agent steering with the adaptivity of multi-agent collaboration.
Experiments on reasoning and personality benchmarks demonstrate that \algname\ consistently outperforms existing single-agent specialization methods and reduces token cost by $20\times$ compared to MAS.

\section*{Limitations}
While \algname\ achieves consistent gains on reasoning quality and efficiency, we acknowledge a few limitations.
First, the proposed \algname\ requires access to the internal hidden states of the backbone LLM and is therefore not directly applicable to black-box API models.
Second, although the codebook is initialized from interpretable text prompts, the learned candidate vectors are latent.
Improving the interpretability of the learned vectors could provide a better understanding of how different candidates contribute to model behavior.
Third, \algname\ composes a query-level persona and keeps it fixed throughout generation. More fine-grained routing at the token or reasoning-step level may enable additional adaptivity, although it could also introduce extra computational overhead.
\bibliography{main}
\newpage

\appendix
\section*{Appendix}

\begin{figure*}[t]
    \centering
    \includegraphics[width=\linewidth, trim = 20 0 10 0]{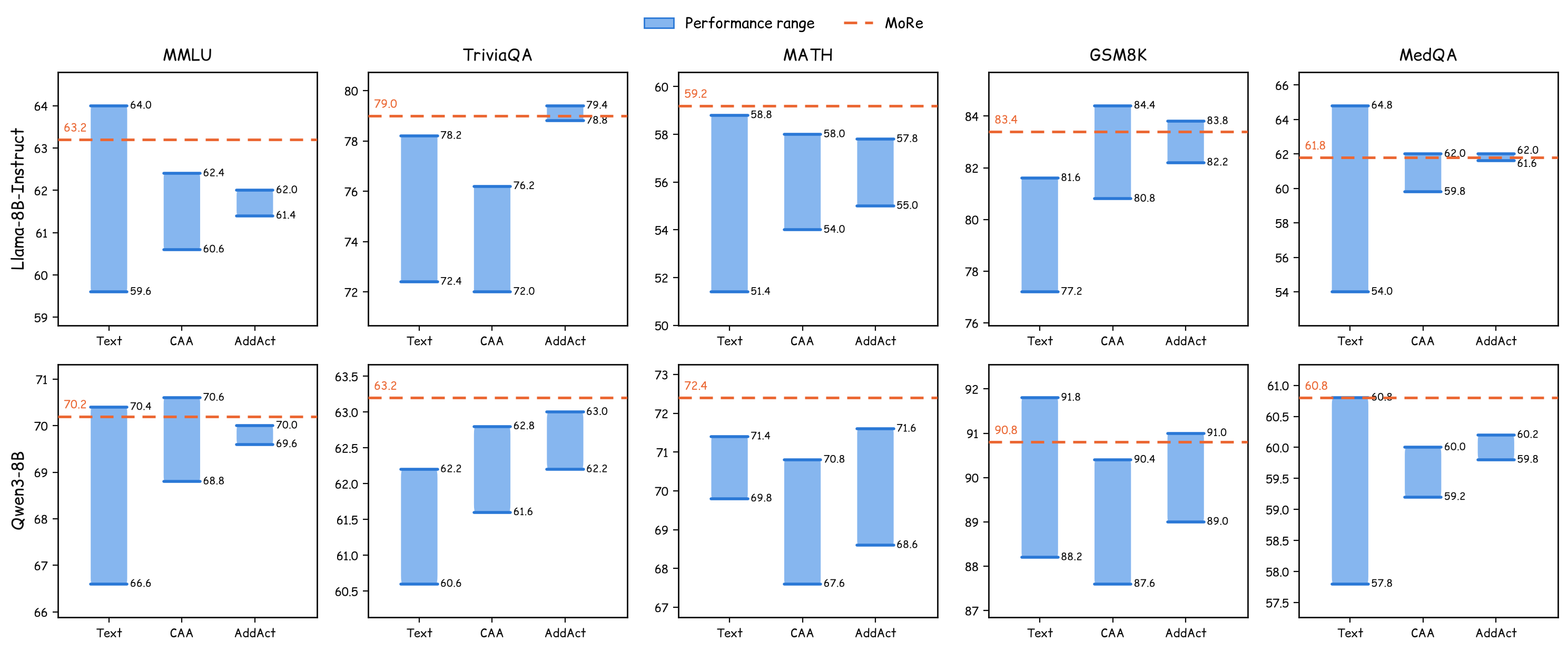}
    \caption{Per-role performance ranges compared with \algname. For each method, the blue bar spans the lowest to the highest score achieved across different roles, while the red dashed line denotes the performance of \algname.}
    \label{fig:app_per_role}
\end{figure*}

\begin{figure*}[t]
    \centering
    \includegraphics[width=\linewidth, trim = 20 0 10 0]{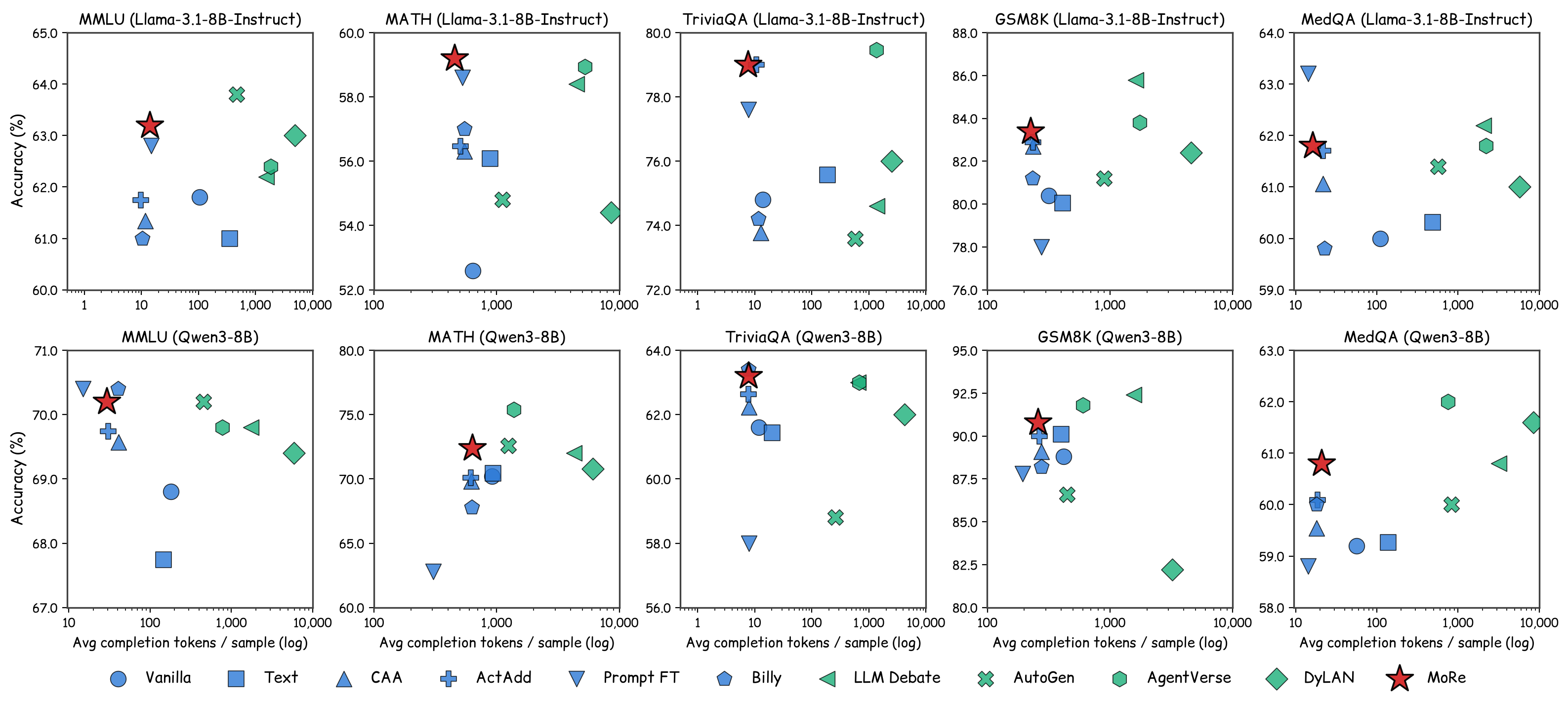}
    \caption{Performance vs \#token on different models and datasets. Left upper corner represents the Pareto front with the best performance and minimal inference cost.}
    \label{fig:app_perf_token}
\end{figure*}

\section{Additional Experiment Results}
\subsection{Per-role Performance}\label{app:study}
To better understand the effect of roles on performance, we report the per-role results in Tables~\ref{tab:app_text}--\ref{tab:app_actadd} and summarize their performance in Figure~\ref{fig:app_per_role}.
\begin{table}
\small
\centering
\caption{Per-role performance on \texttt{Text}.}
\label{tab:app_text}
\setlength{\tabcolsep}{1pt}
\resizebox{\linewidth}{!}{%
\begin{tabular}{lcccccc}
\toprule
\textbf{Role} & \textbf{MMLU} & \textbf{TriviaQA} & \textbf{MATH} & \textbf{GSM8K} & \textbf{MedQA} & \textbf{AvgAcc} \\ \midrule
\multicolumn{7}{c}{\texttt{Llama-8B-Instruct}}  \\ \midrule
Math.      & 64.00 & 73.60 & 58.40 & 81.40 & 61.00 & 67.68 \\
SDE        & 59.60 & 75.60 & 55.00 & 79.60 & 54.00 & 64.76 \\
DataSci.  & 60.20 & 75.40 & 56.40 & 81.60 & 61.40 & 67.00 \\
Logic.   & 61.00 & 72.40 & 54.40 & 77.20 & 55.80 & 64.16 \\
Teacher    & 61.80 & 78.20 & 58.20 & 79.80 & 61.00 & 67.80 \\
Skeptic    & 60.20 & 77.60 & 51.40 & 80.40 & 64.20 & 66.76 \\
Doctor     & 60.20 & 76.20 & 58.80 & 80.40 & 64.80 & 68.08 \\ \midrule
\multicolumn{7}{c}{\texttt{Qwen3-8B}} \\ \midrule
Math.      & 70.40 & 61.60 & 70.40 & 91.80 & 59.00 & 70.64 \\
SDE        & 67.40 & 61.80 & 71.40 & 91.00 & 59.80 & 70.28 \\
DataSci.  & 66.60 & 61.00 & 69.80 & 90.20 & 58.40 & 69.20 \\
Logic.   & 67.20 & 61.80 & 70.60 & 89.80 & 57.80 & 69.44 \\
Teacher    & 67.00 & 60.60 & 69.80 & 90.20 & 60.20 & 69.56 \\
Skeptic    & 68.00 & 62.20 & 70.00 & 89.60 & 58.80 & 69.72 \\
Doctor     & 67.60 & 61.00 & 71.00 & 88.20 & 60.80 & 69.72 \\ \bottomrule
\end{tabular}
}
\vspace{10pt}

\caption{Per-role performance on \texttt{CAA}.}
\label{tab:app_caa}
\setlength{\tabcolsep}{1pt}
\resizebox{\linewidth}{!}{%
\begin{tabular}{lcccccc}
\toprule
\textbf{Role} & \textbf{MMLU} & \textbf{TriviaQA} & \textbf{MATH} & \textbf{GSM8K} & \textbf{MedQA} & \textbf{AvgAcc} \\ \midrule
\multicolumn{7}{c}{\texttt{Llama-8B-Instruct}}  \\ \midrule
Math.      & 60.60 & 74.40 & 58.00 & 84.40 & 60.40 & 67.56 \\
SDE        & 60.60 & 72.00 & 56.20 & 80.80 & 61.20 & 66.16 \\
DataSci.  & 62.00 & 76.20 & 56.80 & 80.80 & 59.80 & 67.12 \\
Logic.   & 61.00 & 74.60 & 54.00 & 83.60 & 61.40 & 66.92 \\
Teacher    & 61.60 & 72.60 & 56.80 & 82.00 & 61.80 & 66.96 \\
Skeptic    & 62.40 & 73.60 & 55.60 & 82.80 & 62.00 & 67.28 \\
Doctor     & 61.20 & 73.00 & 56.80 & 84.40 & 60.80 & 67.24 \\ \midrule
\multicolumn{7}{c}{\texttt{Qwen3-8B}} \\ \midrule
Math.      & 70.40 & 62.00 & 69.80 & 90.40 & 59.60 & 70.44 \\
SDE        & 68.80 & 62.40 & 70.40 & 89.80 & 59.60 & 70.20 \\
DataSci.  & 69.40 & 61.60 & 67.60 & 88.00 & 59.40 & 69.20 \\
Logic.   & 69.60 & 62.20 & 70.20 & 89.40 & 59.20 & 70.12 \\
Teacher    & 69.00 & 62.80 & 70.20 & 88.80 & 60.00 & 70.16 \\
Skeptic    & 69.20 & 62.20 & 70.80 & 89.60 & 59.60 & 70.28 \\
Doctor     & 70.60 & 62.40 & 69.80 & 87.60 & 59.40 & 69.96 \\ \bottomrule
\end{tabular}
}
\vspace{10pt}

\caption{Per-role performance on \texttt{ActAdd}.}
\label{tab:app_actadd}
\setlength{\tabcolsep}{1pt}
\resizebox{\linewidth}{!}{%
\begin{tabular}{lcccccc}
\toprule
\textbf{Role} & \textbf{MMLU} & \textbf{TriviaQA} & \textbf{MATH} & \textbf{GSM8K} & \textbf{MedQA} & \textbf{AvgAcc} \\ \midrule
\multicolumn{7}{c}{\texttt{Llama-8B-Instruct}}  \\ \midrule
Math.      & 61.80 & 79.20 & 55.20 & 82.40 & 61.60 & 68.04 \\
SDE        & 61.60 & 78.80 & 55.00 & 83.40 & 61.60 & 68.08 \\
DataSci.  & 62.00 & 79.40 & 56.60 & 83.40 & 61.60 & 68.60 \\
Logic.   & 61.80 & 78.80 & 57.80 & 82.40 & 62.00 & 68.56 \\
Teacher    & 61.60 & 79.00 & 57.80 & 83.80 & 61.80 & 68.80 \\
Skeptic    & 61.40 & 78.80 & 56.20 & 82.20 & 61.60 & 68.04 \\
Doctor     & 62.00 & 79.00 & 56.60 & 82.60 & 61.80 & 68.40 \\ \midrule
\multicolumn{7}{c}{\texttt{Qwen3-8B}} \\ \midrule
Math.      & 69.60 & 62.20 & 70.80 & 89.00 & 59.80 & 70.28 \\
SDE        & 69.60 & 63.00 & 70.40 & 89.60 & 60.00 & 70.52 \\
DataSci.  & 70.00 & 62.60 & 68.60 & 89.60 & 60.00 & 70.16 \\
Logic.   & 69.80 & 63.00 & 69.40 & 90.40 & 60.20 & 70.56 \\
Teacher    & 69.60 & 62.60 & 69.60 & 91.00 & 60.20 & 70.60 \\
Skeptic    & 70.00 & 62.40 & 71.60 & 90.60 & 60.20 & 70.96 \\
Doctor     & 69.60 & 62.60 & 70.40 & 89.80 & 60.20 & 70.52 \\ \bottomrule
\end{tabular}
}
\end{table}

We first observe that role effectiveness is highly task-dependent.
Some roles exhibit intuitive alignment with particular tasks: for example, the mathematician role performs strongly on mathematical reasoning benchmarks, while the doctor role achieves the best \texttt{Text} performance on MedQA for both backbones.
However, such alignment is not universal.
Roles emphasizing general reasoning strategies, e.g., teacher, skeptic, and logician, can outperform domain-matched roles on several tasks, and the best role often changes across backbone LLMs and steering methods.
These results suggest that roles induce complementary reasoning biases rather than fixed, task-specific capabilities, and that no single predefined role is consistently optimal.

In contrast, \algname\ generally performs near the upper bound of the per-role ranges or even surpasses the best predefined role.
Such improvements can not be solely achieved by selecting a globally strong role, but instead validates that \algname\ benefits from query-dependent selection and composition of complementary candidate steering vectors.
Consequently, \algname\ produces composed specializations that are not restricted to any single predefined role.

\subsection{Balancing Performance--Efficiency}\label{app:balance}

Figure~\ref{fig:app_perf_token} compares model performance against the number of tokens consumed.
Across different models and datasets, \algname\ consistently occupies the upper-left region, indicating a favorable balance between accuracy and inference cost.
For example, \algname\ achieves the best performance on MATH with Llama while using substantially fewer tokens than all MAS baselines, and remains highly competitive on TriviaQA and GSM8K across both backbones.

Compared with single-agent methods, \algname\ consumes a similar number of tokens but generally achieves stronger performance, with particularly clear gains on MATH and GSM8K.
This suggests that query-dependent composition is especially beneficial for tasks involving longer reasoning trajectories and multiple intermediate decisions.
In contrast, fixed textual roles or steering vectors may provide useful but limited specialization.

MAS baselines occasionally achieve higher accuracy, such as on GSM8K, but typically require hundreds or thousands of completion tokens per sample.
In comparison, \algname\ approaches their performance with one to two orders of magnitude lower token cost.
Furthermore, several MAS methods remain less accurate than \algname\ despite their substantially larger inference budgets, showing that additional agent interactions do not necessarily produce proportional performance gains.
Overall, \algname\ combines the efficiency of single-agent inference with performance competitive with substantially more expensive MAS pipelines.

\subsection{On the Steering Effect}
\begin{figure}[t]
    \centering
    \includegraphics[width=\linewidth, trim = 10 0 0 0]{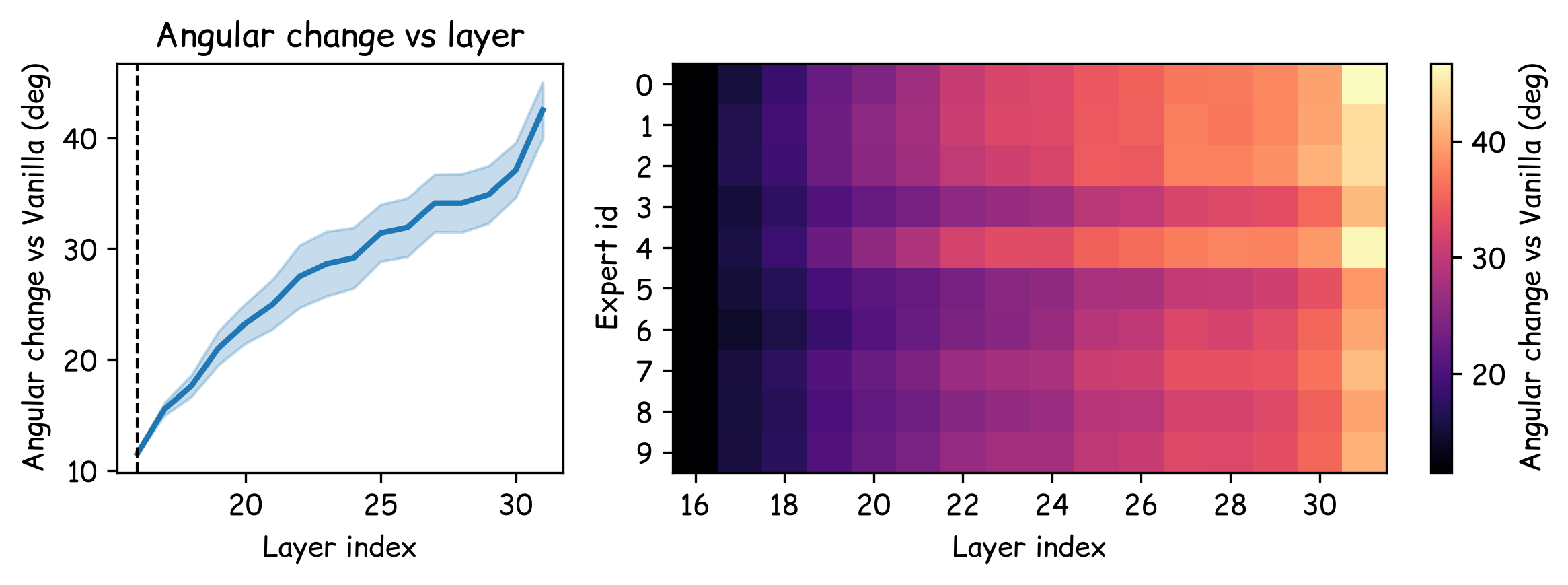}
    \caption{Angular change between last token embeddings with and without steering.}
    \label{fig:app_angular}
\end{figure}

Figure~\ref{fig:app_angular} measures the angular change of the last token embedding with and without steering.
The left figure shows the average angular change and std steered by different expert.
We observe a consistent increase from approximately $12^\circ$ at the steering layer to over $40^\circ$ at the final layer, indicating that the injected perturbation is not attenuated by subsequent computation but is progressively propagated and amplified through the network.

The right figure shows the angular change per expert along the layers.
Although all candidates exhibit the similar increasing trend, their layer-wise trajectories differ noticeably, with the variance across candidates increasing in deeper layers. 
This suggests that the learned candidate vectors induce distinct patterns of representational transformation rather than merely applying perturbations of different magnitudes.
Together, these results show that steering produces persistent and candidate-specific effects that accumulate toward the model output.

\section{Experiment Details}\label{app:exp}

\subsection{Text Role Design}\label{app:persona}
We describe our text role design for \texttt{Text}, \texttt{CAA}, \texttt{Billy}, and the codebook initialization for \algname.
In general, we consider seven persons covering different roles as follows

\begin{tcolorbox}
[left=3pt,right=3pt,top=3pt,bottom=3pt,colback=gray!5!white,colframe=black!75!white,title=Mathematician]
\small
    "You are a Mathematician. Work from precise definitions and exact quantities. Manipulate the problem symbolically, derive each result formally, and justify that every line follows necessarily from the one before."
\end{tcolorbox}

\begin{tcolorbox}
[left=3pt,right=3pt,top=3pt,bottom=3pt,colback=gray!5!white,colframe=black!75!white,title=Software Engineer]
\small
    "You are a Software Engineer. Decompose the task into a concrete procedure, then mentally execute it like code: track the intermediate state, cover boundary and edge cases, and run the inputs back through your answer to test it before finalizing."
\end{tcolorbox}

\begin{tcolorbox}
[left=3pt,right=3pt,top=3pt,bottom=3pt,colback=gray!5!white,colframe=black!75!white,title=Data Scientist]
\small
    "You are a Data Scientist. Reason quantitatively and empirically. Estimate magnitudes, weigh base rates and likelihoods, sanity-check every number against a plausible range, and let the evidence arbitrate between competing answers."
\end{tcolorbox}

\begin{tcolorbox}
[left=3pt,right=3pt,top=3pt,bottom=3pt,colback=gray!5!white,colframe=black!75!white,title=Logician]
\small
    "You are a Logician. List the premises explicitly, then chain strict deductive inferences. Discard any option that forces a contradiction, and surface every hidden assumption and logical fallacy."
\end{tcolorbox}

\begin{tcolorbox}
[left=3pt,right=3pt,top=3pt,bottom=3pt,colback=gray!5!white,colframe=black!75!white,title=Teacher]
\small
    "You are a Teacher. Build intuition in plain language. Motivate why each move makes sense, ground it with a simple analogy or concrete example, and keep the explanation clear enough for a beginner to follow."
\end{tcolorbox}

\begin{tcolorbox}
[left=3pt,right=3pt,top=3pt,bottom=3pt,colback=gray!5!white,colframe=black!75!white,title=Skeptic]
\small
    "You are a Skeptic. Distrust the first answer that comes to mind. Hunt for counterexamples, watch for traps and misleading wording, and re-derive the result from a different angle before you commit to it."
\end{tcolorbox}

\begin{tcolorbox}
[left=3pt,right=3pt,top=3pt,bottom=3pt,colback=gray!5!white,colframe=black!75!white,title=Doctor]
\small
    "You are a Medical Doctor. Gather the salient findings, form a differential of candidate explanations, then rule them out one by one by weighing the evidence for and against each, until the most likely answer remains."
\end{tcolorbox}

\subsection{Reward Design}\label{app:reward}
We utilize GRPO to post train the composer, where the training signal originates from the group-relative advantages of a group of rollouts.
A binary reward based on answer correctness can be problematic in this setting: on samples where rollouts consistently provide correct or wrong answers, the group's reward variance collapses to zero, and the resulting advantage vanishes. 
Therefore, we design soft, task-aligned rewards that (1) preserve a strict correct-over-incorrect ordering, and (2) inject within-group variance so that rollouts sharing the same final-answer verdict are still separable by reasoning quality.
Our designs are as follows
\begin{itemize}[noitemsep, topsep=0pt]
    \item \textbf{Multi-choice questions (MMLU, MedQA).} We assign a stepped reward: $1.0$ for an exact match to the groundtruth, $0.2$ when the model emits a wrong but syntactically valid choice ($A$–$D$) as a format reward, and $0.0$ when no valid choice can be parsed.
    \item \textbf{Free-form QA questions (TriviaQA).} We assign a soft reward: an exact match scores $1.0$; otherwise the reward is the SBERT similarity in $[0,1]$ between the normalized prediction and groundtruth.
    \item \textbf{Long-chain reasoning questions (GSM8K, MATH).} We assign a soft reward. For correct rollouts, we assign a soft reward of $\tau + (1-\tau)\cdot\mathrm{sim}(c, r)$, where $\mathrm{sim}(\cdot,\cdot)$ is the SBERT cosine similarity between the generated chain-of-thought $c$ and the reference rationale $r$. 
    For incorrect rollouts, we assign a continuous partial-credit score $\min\big(\tau,; \lambda_1\cdot\text{cov} + \lambda_2\cdot\mathrm{sim}\big)$, where $\text{cov}$ is the fraction of the reference's intermediate numeric values that also appear in the rollout's reasoning.
\end{itemize}

\subsection{Baseline Methods}\label{app:baseline}
We consider the following single LLM baselines
\begin{itemize}[noitemsep, topsep=0pt]
    \item \texttt{Vanilla} utilizes backbone LLM without any specialization.
    \item \texttt{Text} prepends text persona before the query.
    \item \texttt{CAA}~\cite{rimsky2024steering} extracts persona steering vector from contrastive pairs.
    \item \texttt{ActAdd}~\cite{turner2024activation} extracts token-wise steering vector from contrastive pairs.
    \item \texttt{Prompt FT}~\cite{lester2021power} fine-tunes the soft prompt tokens.
    \item \texttt{Billy}~\cite{pai2026billy} merges multiple contrastive vectors into a single averaged steering vector.
    \item \texttt{NPTI}~\cite{deng2025neuron} modulates neurons for each persona traits in the FFN activations.
    \item \texttt{IRIS}~\cite{wei2026beyond} extends NPTI by retrieving topic-specific neuron sets for adaptive neuron modulation.
\end{itemize}
We also consider the following MAS baselines
\begin{itemize}[noitemsep, topsep=0pt]
    \item \texttt{LLM Debate}~\cite{du2023improving} orchestrates multi-agent to iteratively debate each other's reasoning to reach a consensus.
    \item \texttt{AutoGen}~\cite{wu2023autogen} enables multi-agent collaboration by assigning distinct roles and facilitating multi-turn conversations.
    \item \texttt{AgentVerse}~\cite{chen2024agentverse} coordinates multi-agent via expert recruitment, joint brainstorming, and voting.
    \item \texttt{DyLAN}~\cite{liu2024dynamic} dynamically constructs an agent network that evaluates contributions and routes information among specialized agents across multiple reasoning layers.
\end{itemize}

\subsection{Datasets}\label{app:dataset}
We evaluate on five reasoning benchmarks spanning general knowledge (\textbf{MMLU}, \textbf{TriviaQA}), mathematics (\textbf{MATH},\textbf{GSM8K}), and domain expertise (\textbf{MedQA}), including
\begin{itemize}[noitemsep, topsep=0pt]
    \item \textbf{MMLU}~\cite{hendrycks2020measuring} consists of multiple-choice questions across 57 subjects for assessing general knowledge and problem-solving abilities.
    \item \textbf{GSM8K}~\cite{cobbe2021training} covers grade-school math word problems that require multi-step reasoning.
    \item \textbf{MATH}~\cite{hendrycks2021measuring} covers mathematics problems ranging from grade-school difficulty to advanced competition level.
    \item \textbf{TriviaQA}~\cite{joshi2017triviaqa} covers open-domain factoid question answering and reading comprehension questions.
    \item \textbf{MedQA}~\cite{jin2021disease} consists of multiple-choice questions collected from professional medical board exams.
\end{itemize}

Besides, we adopt \textbf{PersonalityBench}~\cite{deng2025neuron} to evaluate whether LLM responses exhibit specified personality traits.
It is grounded in the Big Five personality model, covering \textbf{O}penness, \textbf{C}onscientiousness, \textbf{E}xtraversion, \textbf{A}greeableness, and \textbf{N}euroticism, including the opposing aspects of each trait.
\textbf{PersonalityBench} utilizes open-ended situational questions derived from real-world behaviors, requiring models to express the target personality through natural-language responses. 
We evaluate on the evaluation set containing 90 questions per personality, and employ an LLM-as-a-judge to assess the extent to which each response reflects the intended personality.

\section{More on Related Works}
We provide more related works on mixture of experts and pre-trained foundation models.

\paragraph{Mixture of Experts}

Mixture-of-Experts (MoE) provides a general framework for conditional computation by dynamically routing different inputs to specialized expert modules~\cite{shazeer2017outrageously,lepikhin2020gshard}.
Modern sparse MoE architectures substantially scale this principle by activating only a small subset of experts for each input~\cite{fedus2022switch,du2022glam,ai2025resmoe,lin2026mixture}.
This paradigm has become increasingly prevalent in LLMs, where models such as Mixtral~\cite{jiang2024mixtral} and DeepSeekMoE~\cite{dai2024deepseekmoe} increase model capacity while maintaining a relatively small number of activated parameters.
In particular, recent designs increasingly emphasize fine-grained expert specialization and flexible expert composition, which decomposes experts into finer-grained modules and introduces shared experts to reduce redundancy among routed experts~\cite{dai2024deepseekmoe}.
Beyond full-parameter experts, the MoE principle has also been extended to parameter-efficient specialization, where lightweight LoRA adapters serve as experts and are dynamically selected or composed according to the input~\cite{liu2024moe,li2024mixlora,zeng2025hierarchical,zeng2026s}.
\algname\ extends the principle of MoE from \emph{parameter-space computation} to \emph{activation-space behavioral specialization}, enabling dynamic multi-perspective composition with minimal additional inference overhead.

\paragraph{Pre-trained Foundational Models}
Foundation models have reshaped the field by pretraining models on vast amounts of web-scale data~\cite{brown2020language,bommasani2021opportunities,achiam2023gpt,grattafiori2024llama}.
Large Language Models (LLMs) built upon this paradigm have demonstrated remarkable performance across diverse tasks including natural language understanding~\cite{comanici2025gemini,yang2025qwen3,wei2026inference}, knowledge-intensive question answering~\cite{achiam2023gpt,zeng2026harnessing,cui2026adafuse,lin2026alert}, multi-modal reasoning~\cite{zeng2026subspace,zeng2025interformer,liang2025external}, and domain-specific problem solving~\cite{wei2022chain,guo2025deepseek, li2026flow, zhang2026improving}.
Beyond general-purpose capabilities, an emerging line of research seeks to \emph{specialize} pretrained LLMs toward particular domains, behaviors, personalities, or reasoning styles~\cite{wei2026agentic,ning2026code}.
Such specialization can be induced through parameter-efficient adaptation, including prompt tuning and lightweight parameter updates~\cite{lester2021power,hu2021lora}, or through role-conditioned prompting that instructs LLMs to behave as particular experts or personas~\cite{shanahan2023role,shao2023character,chen2024from}.
More recently, representation-level approaches directly manipulate internal activations to induce targeted behaviors without modifying the backbone parameters~\cite{li2023inference,turner2024activation,rimsky2024steering}.
These techniques have been further extended to persona and behavioral control through learned steering vectors, neuron-level interventions, and context-dependent specialization~\cite{cao2024personalized,chen2025persona,deng2025neuron,sun2025personality,wei2026beyond}.
Together, these studies demonstrate that pretrained foundation models contain rich latent capabilities that can be selectively elicited through lightweight specialization.
However, most existing specialized-agent approaches rely on a predefined or individually selected specialization, motivating methods that can dynamically compose multiple complementary specializations according to the input query.

\section{Potential Risks}
\algname\ modifies the internal representations of LLMs to induce query-dependent specialization. 
Although our experiments focus on established reasoning and personality benchmarks, the learned steering vectors may amplify undesirable biases or produce unintended behavioral changes on inputs outside the training distribution.

Besides, the personality evaluation relies on LLM-as-a-judge, which may inherit biases, preferences, or inconsistencies from the evaluator model.
We mitigate this by following the standard evaluation protocol of PersonalityBench and applying the same scoring criteria to all methods.

\section{Use Or Create Scientific Artifacts}
Our work builds upon publicly available datasets, pre-trained language models, and existing specialization and multi-agent baselines. We do not collect new data from human participants.

\subsection{Cite Creators Of Artifacts}
All external artifacts are properly credited to their original publications and repositories.

The benchmarks used in this work, including \textbf{MMLU}~\cite{hendrycks2020measuring}, \textbf{TriviaQA}~\cite{joshi2017triviaqa}, \textbf{MATH}~\cite{hendrycks2021measuring}, \textbf{GSM8K}~\cite{cobbe2021training}), \textbf{MedQA}~\cite{jin2021disease} and \textbf{PersonalityBench}~\cite{deng2025neuron}, are credited to their respective authors.

The backbones used in this work, including \texttt{Llama3.1-8B-Instruct}~\cite{grattafiori2024llama} and \texttt{Qwen3-8B}~\cite{yang2025qwen3}, are referenced to its official publication.

We additionally cite the original publications for all single-agent and
multi-agent baselines, including \texttt{CAA}~\cite{rimsky2024steering}, \texttt{ActAdd}~\cite{turner2024activation}, \texttt{Prompt FT}~\cite{lester2021power},
\texttt{Billy}~\cite{pai2026billy}, \texttt{NPTI}~\cite{deng2025neuron}, \texttt{IRIS}~\cite{wei2026beyond}, \texttt{LLM Debate}~\cite{du2023improving}, \texttt{AutoGen}~\cite{wu2023autogen}, \texttt{AgentVerse}~\cite{chen2024agentverse}, and \texttt{DyLAN}~\cite{liu2024dynamic}.

\subsection{Discuss The License For Artifacts}
We comply with the licenses of all artifacts used in this work. The datasets and backbone models are available for research purposes under their respective licenses.

\subsection{Data Contains Personally Identifying Info Or Offensive Content}
We do not collect or introduce new personally identifying information.
The reasoning benchmarks mainly contain general knowledge, mathematical, and professional examination questions. 
PersonalityBench contains synthetically constructed situational questions based on the Big Five personality framework. 
These questions may describe interpersonal conflicts, emotional states, or other sensitive situations necessary for evaluating personality expression.
To our knowledge, they do not contain private information about identifiable individuals.

\subsection{Documentation Of Artifacts}
Appendix~\ref{app:exp} documents the evaluated backbone models,
datasets, baselines, evaluation protocols, and hardware.
Dataset statistics are summarized in Appendix~\ref{app:stat}. 
These materials specify how each artifact is used and support reproducibility of the experiments.

\section{Statistics For Data}\label{app:stat}
We summarize the dataset statistics as follows

\subsection{MMLU}
\begin{itemize}[noitemsep, topsep=0pt]
    \item Domain: Multiple-choice knowledge and reasoning questions covering
    57 subjects in the humanities, social sciences, natural sciences, and
    professional domains. Each question has four candidate answers.
    \item Size: Approximately 16,000 questions.
    \item Metric: exact match
\end{itemize}

\subsection{TriviaQA}
\begin{itemize}[noitemsep, topsep=0pt]
    \item Domain: Open-domain factoid question answering covering diverse topics such as history, geography, science, entertainment, and popular culture. Questions require free-form textual answers.
    \item Size: Approximately 95,000 question--answer pairs.
    \item Metric: accuracy given by LLM-as-a-judge.
\end{itemize}

\subsection{MATH}

\begin{itemize}[noitemsep, topsep=0pt]
    \item Domain: Competition-level mathematical problem solving across algebra, counting and probability, geometry, intermediate algebra, number theory, prealgebra, and precalculus. Each problem requires a free-form solution and final answer.
    \item Size: 7,500 training problems and 5,000 test problems.
    \item Metric: accuracy given by LLM-as-a-judge.
\end{itemize}

\subsection{GSM8K}
\begin{itemize}[noitemsep, topsep=0pt]
    \item Domain: Multi-step grade-school mathematical word problems that require arithmetic reasoning and the generation of a numerical answer.
    \item Size: 7,473 training problems and 1,319 test problems.
    \item Metric: accuracy given by LLM-as-a-judge.
\end{itemize}

\subsection{MedQA}
\begin{itemize}[noitemsep, topsep=0pt]
    \item Domain: Multiple-choice medical question answering derived from
    professional medical licensing examinations. We use the English subset,
    which primarily contains questions based on the United States Medical
    Licensing Examination.
    \item Size: 12,723 questions across the official training, development, and test splits.
    \item Metric: exact match
\end{itemize}

\subsection{PersonalityBench}

\begin{itemize}[noitemsep, topsep=0pt]
    \item Domain: Open-ended situational question answering for evaluating
    personality expression.
    \item Size: 180,000 training instances, with approximately 36,000 instances for each personality trait. Its evaluation set contains approximately 90 situational questions for
    each trait.
    \item Metric: LLM-as-a-judge to score each response on a scale from 1 to 5. We report the sum of the scores for the high and low poles of each trait, as well as the standard deviation across traits.
\end{itemize}

\section{Computational Experiments}\label{sec:app-exp}
All computational experiments in this work are fully reproducible, with details provided in Section~\ref{sec:setup}.

\subsection{Model Size And Budget}
We evaluate \algname\ on two open-weight instruction-following LLMs:
\begin{itemize}[noitemsep, topsep=0pt]
    \item \texttt{Llama-3.1-8B-Instruct}: approximately 8 billion
    parameters.
    \item \texttt{Qwen3-8B}: approximately 8 billion parameters.
\end{itemize}

The parameters of both backbone LLMs remain frozen throughout training. 
Only the codebook of candidate steering vectors and the lightweight query-aware composer are optimized. 
The resulting number of parameters 0.04M for codebook and 1.32M for router, sum of which constitutes 0.017\% of the backbone model.

All experiments are conducted on 8 NVIDIA A100 GPUs with 40GB memory.

\subsection{Experimental Setup And Hyper-params}
We describe experimental settings in Section~\ref{sec:setup}.
Our default configuration for key hyperparameters studied includes:
\begin{itemize}[noitemsep, topsep=0pt]
    \item Steering strength $\alpha=0.2$.
    \item Codebook size $N=10$.
    \item Sparse routing Top-$K$ with $K=3$.
    \item Load-balancing weight $\lambda_{\mathrm{lb}}=0.1$.
\end{itemize}

We study the sensitivity to the following hyperparameters in Section~\ref{sec:hyper}, including Steering strength $\alpha\in\{0.1,0.2,0.4,0.6,0.8\}$; Codebook size $N\in\{4,7,10,13,16\}$; Router sparsity $K\in\{1,3,5,7,10\}$; and Load-balancing weight $\lambda_{\mathrm{lb}}\in\{0,10^{-3},10^{-2},10^{-1},1\}$.

\subsection{Descriptive Statistics}
For the five reasoning benchmarks, we report accuracy on each dataset, average accuracy across datasets, and average rank across methods.
For PersonalityBench, we report the LLM-judge score for each Big Five trait, the average score across traits, and the standard deviation between the high- and low-pole evaluations.

\subsection{Parameters For Packages}
The existing packages used are specified as follows, including \texttt{Python}: 3.10.20, \texttt{PyTorch}: 2.9.1 (CUDA 12.8), \texttt{transformers}: 5.10.2, \texttt{datasets}: 5.0.0, \texttt{sentence-transformers}: 5.5.1 and \texttt{vllm} 0.6.2.

\section{AI Assistants In Research Or Writing}
In this paper, AI assistant tool is used to edit and improve the quality of the text, including checking the spelling, grammar, punctuation and clarity.

\end{document}